\documentclass[journal]{IEEEtran}

\usepackage{graphicx}
\usepackage{cite}
\usepackage{comment}
\usepackage{amsmath,amssymb,amsfonts}
\usepackage{booktabs} 
\usepackage{dblfloatfix}

\newcommand{{\VO}}{$\textrm{VO}_2$}
\newcommand{\vhtable}{\rule{0pt}{10pt}}% defines the separation and can be different if you so desire for each different row.

\usepackage{xcolor}
\definecolor{R1}{rgb}{1 0 0}
\definecolor{R2}{rgb}{0 0 1}
\definecolor{R12}{rgb}{0.33 0.08 0.7}
\newcommand{\rev}[1]{#1}
\newcommand{\srev}[1]{}
\usepackage[normalem]{ulem}

\begin{document}

\title{\rev{Dynamically Reconfigurable Microwave Circuits Leveraging Abrupt Phase-Change Material}}

\author{David Connelly,~\IEEEmembership{Student Member,~IEEE}, Jonathan Chisum, ~\IEEEmembership{Senior Member,~IEEE}%
\thanks{The authors are with the Department of Electrical Engineering, University of Notre Dame, Notre Dame, IN, 46556 USA (e-mail: dconnel7@nd.edu, jchisum@nd.edu).}
}
\maketitle

\begin{abstract}
This article \rev{proposes a concept for dynamically} reconfigurable distributed microwave circuits \rev{by leveraging the abrupt conductivity transition in phase-change materials (PCM)}. Metallic \rev{surface-}inclusions (\boldmath$\ll$\boldmath$\lambda$) are embedded in the PCM film---vanadium dioxide ({\VO})---to provide low-loss and reconfigurability. To validate this concept, a variety of co-planar waveguide transmission lines are designed and fabricated with metallic \rev{surface-}inclusions in {\VO} films, and the lines' performance are characterized up to 50\,GHz. The measured results are used to develop a transmission-line based model and an equivalent circuit model of the {\VO} with \rev{surface-}inclusions to aid in the rapid design of new structures. Additionally, an electromagnetic model was developed and indicates that loss can be close to that of conventional metallic distributed circuits with 100 to 200\,nm thick {\VO} films. \rev{With these thicker films, two practical realizations of this concept were designed and simulated: a tunable dipole from 2.13 to 9.07\,GHz and a tunable triple-stub matching network from 5 to 40 GHz with high $|\Gamma|$.} Therefore, the proposed method \srev{is} \rev{appears} viable for the realization of arbitrary programmable distributed circuits and antennas in the microwave and \rev{low-millimeter-wave} bands.
\end{abstract}

% Note that keywords are not normally used for peer review papers.

\IEEEpeerreviewmaketitle

\section{Introduction}

\begin{figure}[t]
\centering
\includegraphics[width=\columnwidth]{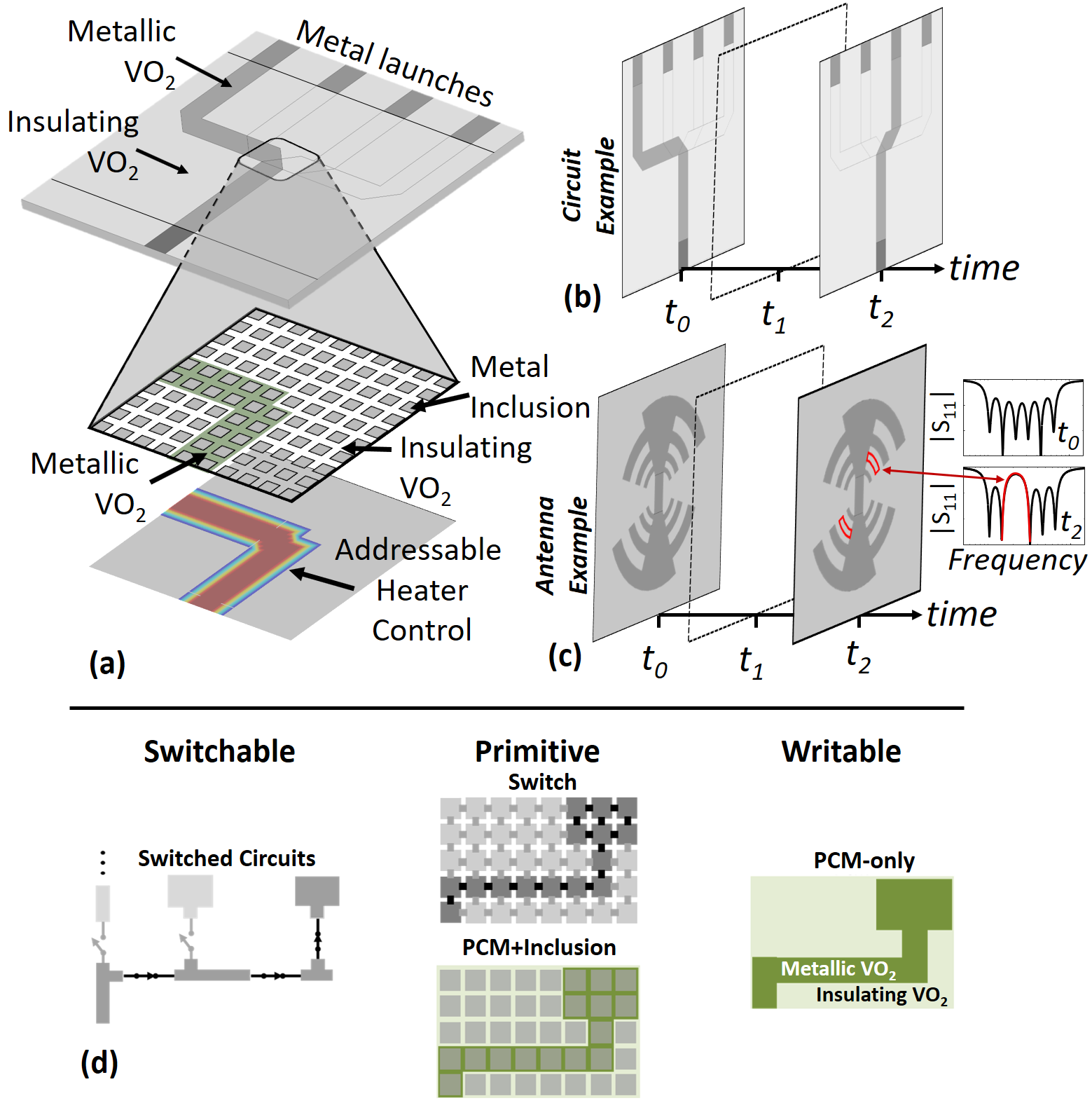}
\caption{(a) CONCEPT: {\VO} film on a microwave substrate is patterned with conductive \rev{surface-}inclusions to reduce loss. Regions of metallic- and insulating-{\VO} are defined (here, by a heater array) to create a distributed circuit or antenna. (b) \textit{Distributed circuit:} event at $t_1$ triggers reconfiguration of a switch matrix from output one to three. (c) \textit{Antenna scenario:} a log-periodic tooth antenna at $t_0$ exhibits a wideband input match (inset black $|S_{11}|$ trace). An interference source is detected at time $t_1$ which triggers reconfiguration and removal of one set of resonant teeth (outlined in red). The input match at $t_2$ exhibits a notch (red arrow) to reduce interference. (d) Reconfigurable RF circuits can be realized along a continuum from coarsely switched functional circuits (e.g. switch matrix and antenna array) to large arrays of programmable, interconnected conductive \rev{surface-}inclusions to fully writable PCMs. This article examines the feasibility of metal \rev{surface-}inclusions in a {\VO} film (``PCM+Inclusion'') to reduce the loss of the ``Writable'' {\VO} PCM-only approach.}
\label{fig:Patchconcept}
\end{figure}

\IEEEPARstart{D}{istributed} microwave circuits and antennas derive their functionality and performance from their geometry with respect to an electromagnetic wavelength. Such geometries are typically defined through a material contrast in order to efficiently guide, confine, radiate or otherwise manipulate electromagnetic energy. At microwave and millimeter-wave frequencies, the material contrast for passive microwave circuits (e.g., microstrip, co-planar waveguide) and antennas (e.g., dipoles, loops, bow-ties, spirals, patches, etc.) is predominantly realized through geometric patterning of metals on insulating substrates. Because metallic geometry is defined at fabrication time, microwave circuits and antennas are not field-reconfigurable. And yet, nearly every other electrical component from a microprocessor to a Field Programmable Gate Array (FPGA) to a switched analog circuit is field-programmable on some level. It is the fundamental link between functionality and geometry which makes the realization of programmable microwave circuits very challenging. In order to realize fully programmable distributed microwave circuits and antennas, there is a need to locally and dynamically set a circuit's conductivity between that of a metal and that of an insulator. \rev{In this article we will present a concept for the dynamic reconfiguration of microwave circuits (including antennas) which leverages the abrupt conductivity transition in phase-change materials (PCM), as depicted in Fig.\,\ref{fig:Patchconcept}(a).}

The family of PCMs, including vanadium dioxide {\VO} \cite{growth,growth2} and germanium telluride (GeTe) \cite{TomerGeTeReconfigCircuits_2018}, has the ability to change its state from insulating to metallic via several mechanisms, including local heating and even localized illumination with high energy (UV) light. PCMs have been used to realize reconfigurable RF circuits and antennas based upon switches \cite{switch,darpa_act_rffpga}, tunable capacitors \cite{VO2tunablecapsfilters} and with them, tunable filters \cite{VO2tunablecapsfilters}. However, these approaches only take advantage of the conductivity transition as a switch mechanism in statically-defined distributed circuits and have stopped short of realizing fully programmable conductor geometry for distributed circuits. The possibility of locally imprinting metallic states across an entire {\VO} film for programmable circuits was first suggested in \cite{opt}, but neither details nor demonstrations have been offered. One of the challenges of such an approach is that the metallic-state conductivity of {\VO} is two orders of magnitude lower than that of a typical metal ($\sigma$\textsubscript{VO2} $\approx$ 10$^5$\,S/m). It is worth noting that NiCr, a material typically used to fabricate microwave resistors \cite{NiCrRes}, has a conductivity on the order of $10^5$\,S/m, so even in the metallic state, {\VO} behaves more like a thin-film resistor than like a thin metallic conductor.

Past efforts to achieve a programmable conducting surface generally fall into two categories: either patterning a PCM film with a spatially programmable heater array as was simulated in \cite{Gerislioglu_VO2AntennaWithMicroheaterArray_2017,patent_pala_2018}, or interconnecting conducting regions with switches. The conducting regions are either traditional microwave structures such as patch antennas \cite{ali_memspatcharray,patent_jackson_2005} or dipole arrays \cite{Chau_GeTeAntennas}, or they are sub-wavelength metallic structures as proposed in \cite{patent_schaffner_2016}. This continuum is depicted in Fig.\,\ref{fig:Patchconcept}(d) with writable PCM-only films on the right and switch-interconnected traditional microwave structures on the left. 

Pure PCM thin films (Fig.\,\ref{fig:Patchconcept}(d): ``Writable'') have insufficient conductivity in the metallic state and thus result in significant loss through microwave structures. For {{\VO}} film thickness and conductivity of 35\,nm and 10$^5$\,S/m, insertion loss is 35--100\,dB/mm across 1--50\,GHz, rendering this approach infeasible for distributed circuits and antennas. However, if high quality ($>$ 10$^5$\,S/m) films of significant thicknesses (on the order of a skin-depth) become possible, then a pure {{\VO}} approach could be considered. The opposite end of the reconfigurable continuum consists of interconnected conductive regions with electrical switches (Fig.\,\ref{fig:Patchconcept}(d): ``Switchable'') and suffers from poor spatial resolution and complex switch routing which interacts electromagnetically with the programmed structure. 

In between these two approaches is a third option (Fig.\,\ref{fig:Patchconcept}(d): ``Primitive: PCM+Inclusion''), where metal is deposited discretely as small ($\ll$ $\lambda$), isolated \rev{surface-}inclusions over a film of PCM material in order to raise the effective conductivity. This concept was proposed in \cite{patent_schaffner_2016} but was not demonstrated or studied in detail. The PCM film can change into a metallic state with either spatially controlled heating (e.g., with a discrete micro-heater array or with a scanned infrared laser) or by raster scanning an ultraviolet laser of sufficient energy level \cite{opt}. By such means, the control mechanism is displaced from the electromagnetic structure minimizing electromagnetic interaction between the two.

Figure\,\ref{fig:Patchconcept}(a) shows the proposed concept in which a phase-change material, here a {\VO} film, containing metallic \rev{surface-}inclusions is placed in close proximity to a thermal control plane---here a heater array---which is used to thermally bias the {\VO} film in specified regions. The result is a region of metallic {\VO} (indicated as green) where the localized temperature is above the transition temperature ($\sim$68$^\circ$C) within the otherwise insulating {\VO} (clear) regions. \rev{Key to the success of such an approach is the abrupt transition of {\VO} conductivity---the insulating to metallic {\VO} conductivity ratio is $10^4$ within approximately 4$^\circ$C\cite{comb,transition_vo2}---requiring only a small variation in control temperature profile to switch adjacent cells on an off. Further, this abrupt transition occurs in less than one microsecond \cite{VO2_SPST}.} This concept can be used to realize \rev{dynamically} programmable distributed circuits (Fig.\,\ref{fig:Patchconcept}(b) and even programmable and reconfigurable antennas (Fig.\,\ref{fig:Patchconcept}(c)). 

Both Fig.\,\ref{fig:Patchconcept}(a) and (b) show an example of a programmable distributed circuit: a single-pole, four-throw (SP4T) switch realized by programming a transmission line between the common and the output port of interest. Figure\,\ref{fig:Patchconcept}(b) shows an hypothetical time-sequence wherein the common port of the distributed SP4T switch is switched from output port 1 at time $t_0$ to output port 3 at time $t_2$ due to a reconfiguration event starting at time $t_1$. While the benefit of such a realization may not be obvious with an SP4T switch, the benefit becomes clear for a larger number of series switches in each path which results in performance degradation due to reflections and switch loss. In contrast, the programmable switch matrix could have low loss regardless of the switch matrix complexity or function (e.g., blocking and non-blocking).

Figure\,\ref{fig:Patchconcept}(c) shows that the same concept can be used to realize programmable antennas as well. Here, a toothed log-periodic antenna is reconfigured at time $t_1$ in response to a detected interference source. A pair of teeth with resonance frequency corresponding to the interference frequency is omitted from the antenna (as in \cite{mruk}) to produce a rejection band in the antenna's response. Such a reconfigurable antenna would be useful for cognitive radios in congested environments and for electronic warfare sensors. These two examples illustrate the vast number of possible circuits and antennas one could realize with such a system.

Unfortunately, the necessary integration of PCM and metallic \rev{surface-}inclusions in order to overcome significant RF loss introduces a trade-off between reducing loss in the metallic state (ON-state) and increasing isolation in the insulating state (OFF-state). Capacitive coupling between metal \rev{surface-}inclusions in regions of insulating {\VO} results in reduced isolation that degrades with increasing frequency. Effectively, the boundary between conducting and insulating regions of the reconfigurable surface becomes poorly defined. To achieve a low-loss distributed circuit, the percentage of metal must be high and yet, to achieve sufficient reconfigurability, the percentage of metal must be small enough so as to present a well-defined and finely discretized reconfigurable current path.

The purpose of this article is to provide the first demonstration of microwave performance for a {\VO} structure with metallic \rev{surface-}inclusions \rev{here after also referred to as inclusions} in order to i) determine the feasibility of such a structure for \srev{low-loss} distributed circuits and antennas, ii) explore the trade-off between loss and isolation versus inclusion geometry, \srev{and }iii) to develop models for design and simulation purposes\rev{, and iv) to demonstrate reconfigurable circuits with this concept}. To this end, we designed and fabricated unit-cells comprising various percentages of {\VO} film and conductive inclusions, and we performed measurements from 10\,MHz to 50\,GHz. Based on these measurements, we develop models for circuit, transmission-line, and full-wave electromagnetic simulations. The electromagnetic models are based upon low-frequency material parameter extraction and achieve good agreement with measurement except for minimal tuning of these parameters. Parasitic circuit models are developed which explain the difference between ideal distributed circuit elements (e.g., stubs) and the same structures realized with the proposed {\VO} film with metallic inclusions. In order to simplify the structure for fabrication and measurement, we have reduced the two-dimensional (2D) reconfigurable surface to a one-dimensional (1D) reconfigurable transmission line. We note that the models developed from measured 1D structures are still relevant to 2D structures—--specifically, full-wave EM model parameters are identical if the {\VO} film quality is the same.

Section \ref{sec:fabrication} presents the fabricated test structures and section \ref{sec:measurement} presents measurements, calibration, de-embedding, \srev{and} unit-cell model extraction\rev{, dispersion analysis, and maximum operating frequency}. Section \ref{sec:implementation} extends the present work with a discussion of practical requirements on film \rev{and metallization} thickness for the realization of high-performance structures\rev{, as well as a preliminary yet quantitative thermal control study. Section \ref{sec:application} presents simulated demonstrations of a tunable dipole and triple-stub matching network, and ends with a discussion of a two-dimensional reconfigurable switch matrix.} Final conclusions and future work are presented in section \ref{sec:conclusion}.

\section{Sample Fabrication} \label{sec:fabrication}

\begin{figure}[t]
\centering
\includegraphics[width=\columnwidth]{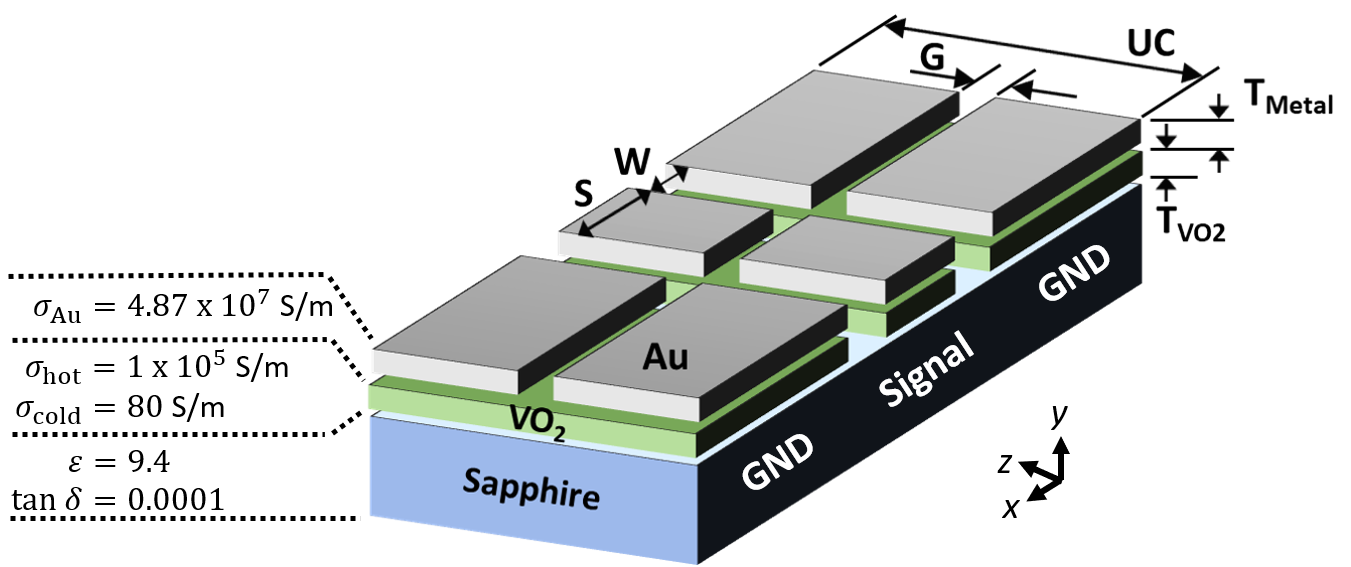}
\caption{Unit-cell of a CPW transmission line comprising a {\VO} film (green layer) with gold inclusions (gray layer) on a sapphire substrate (blue layer). The power flow is along the z--axis and the CPW cross-section is in the x--y plane. Fabricated unit-cell parameters (G, UC, S and W) are defined in Table \ref{tab:designparam}. The {\VO} conductivities are the final values used to achieve the best fit between electromagnetic models and measurements. Substrate thickness was measured to be 452\,$\mu$m.}
\label{fig:UnitcellParameters}
%\vspace{-0.5cm}
\end{figure}

\begin{figure}[!ht]
\centering
\includegraphics[width=\columnwidth]{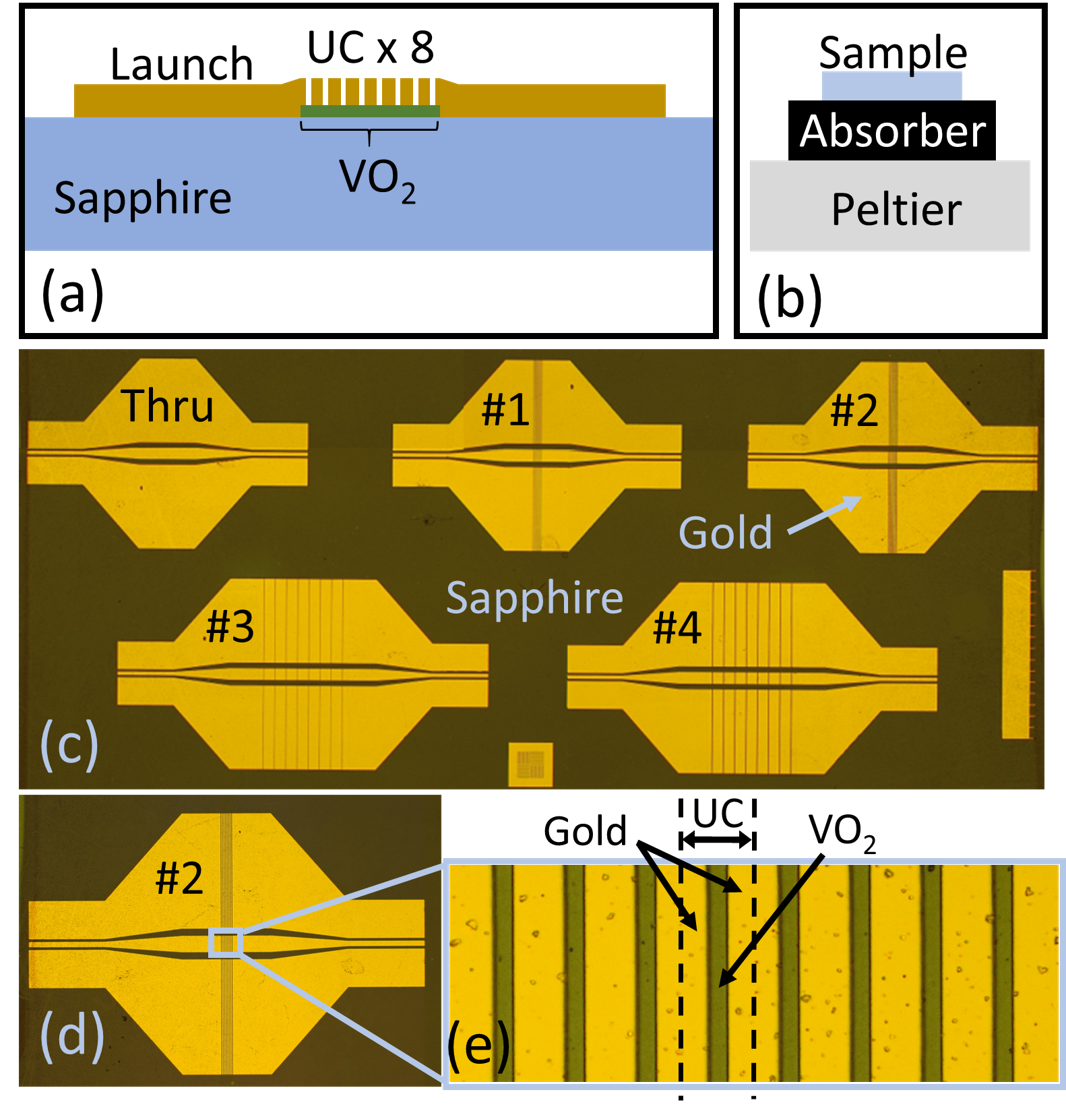}
\caption{(a) Side-view of a representative structure that comprises {\VO} underneath all 8 UC but not under the launches. (b) The sample was measured using a carbon-impregnated AlN absorber to remove from the waveguides any possible ground-plane effects, and the sample was heated using a thermo-electric Peltier heater. (c) An overview photograph of the fabricated structures including a THRU standard for de-embedding launches and four structures with varying UC length and gap size. Structures labeled with sample number \#1\,--\,\#4 have dimensions listed in Table\,\ref{tab:designparam}. Close-ups of a structure identifying the unit-cell are provided in (d) and (e).} 
\label{fig:sample}
%\vspace{-0.5cm}
\end{figure}

To investigate the trade-off between ON-state loss and OFF-state isolation as a function of unit-cell parameters (Fig.\,\ref{fig:UnitcellParameters}) four co-planar waveguides (CPW) with 8 unit-cells were designed and parameterized by unit-cell length (UC) and unit-cell gap (G) as given in Table \ref{tab:designparam} and shown in Fig.\,\ref{fig:sample}. Gap sizes of 1 and 2$\,\mu$m were chosen to facilitate the fabrication of these structures with photolithography but smaller gaps could be fabricated with more advanced processes (e.g., electron-beam lithography). Unit-cell lengths of 10\,$\mu$m and 100\,$\mu$m were chosen to vary the ratio of metal to {\VO} in each unit-cell. The 100\,$\mu$m unit-cell is expected to exhibit lower on-state insertion loss per unit length at the expense of reduced spatial resolution.

\begin{table}[tb]
\centering
\caption{Fabricated Parameters for Each Structure (Fig.\,\ref{fig:UnitcellParameters}).}
\begin{tabular}{@{}ccccc@{}}
    \toprule
        Structure & G & UC & W / S & T\textsubscript{VO2}\ /\ T\textsubscript{METAL} \\
         & ($\mu$m) & ($\mu$m) & ($\mu$m) & (nm)\\
    \midrule
    \vhtable    1  & 1  & 10  & 45 / 106.5 & 17 / 200 \\
    \vhtable    2  & 2  & 10  & 45 / 106.5 & 17 / 200 \\
    \vhtable    3  & 1  & 100 & 45 / 106.5 & 17 / 200 \\
    \vhtable    4  & 2  & 100 & 45 / 106.5 & 17 / 200 \\
    \bottomrule
\end{tabular}
%\vspace{0.1in}
\label{tab:designparam}
\end{table}

A THRU standard to de-embed the launches was also included. Due to the size of the 50\,GHz RF probes used in this work (150\,$\mu$m pitch), a transition from CPW geometry (S,\,W) = (36,\,16)\,$\mu$m to (108,\,45)\,$\mu$m was necessary to measure larger unit-cell geometries. Since the CPW probe pads and the larger unit-cell cross-sections have a characteristic impedance of 50\,$\Omega$, a matching network is not needed; instead, we require a mode transition which maintains a 50\,$\Omega$ environment between the mode of the smaller and larger geometries. ADS Momentum was used to confirm that the mode transition maintained $|S_{11}|< -$20\,dB across the band (0.01--50\,GHz). 

All experimental work shown here was done using a nominally 20\,nm thick {\VO} sample on r-plane sapphire grown using the method of \cite{comb} at Pennsylvania State University. Waveguides were patterned on this sample using the University of Notre Dame Nanofabrication Facility. The sample was patterned by photolithography using nLOF 2020 photoresist and the oxide etched in a Reactive Ion Etcher (RIE) using tetrafluoromethane (CF4) as the etching agent. Lift-off was performed with Remover PG at 80$^\circ$C after electron-beam evaporating 20\,nm of palladium (seed layer) and 180\,nm of gold (Au). Palladium was chosen instead of other seed layers, such as titanium, because it does not scavenge the oxygen in the film (which would increase conductivity in the OFF-state). The resulting structures are shown in Fig.\,\ref{fig:sample}. The oxide was etched everywhere except underneath and in-between the metal inclusions (Fig.\,\ref{fig:sample}a). There is no oxide underneath the launches. The film was etched to eliminate the need to include a micrometer heater array design. This way, the entire sample could be heated or cooled to induce the phase change of {\VO} in the un-etched regions.

Typical cleanroom processes, when performed for extended periods of time, seem to affect the quality (stoichiometry), thickness, and conductivity of {\VO} \cite{percy_PhDThesis}. The potentially detrimental processes utilized on the sample presented in this paper include cleaning with RF oxygen plasma, rinses with photoresist removal agents and de-ionized water, and heating (max. 110\,$^\circ$C). However, the processing time required for this sample was very brief: seconds or minutes instead of the hours taken in the investigation in \cite{percy_PhDThesis}. Additionally, the film's DC conductivity in the ON and OFF state was measured before and after sample fabrication using a van der Pauw method. The results are $\sigma_{\textrm{ON}} = $ 3.61$\times$10$^5$\,S/m and $\sigma_{\textrm{OFF}} = $ 95.3\,S/m pre-fabrication, which is a $\Delta\sigma = $ 3.58 orders of magnitude. Post-fabrication measurements yield $\sigma_{\textrm{ON}} = $ 1.65$\times$10$^5$\,S/m and $\sigma_{\textrm{OFF}} = $ 188\,S/m, which is a $\Delta\sigma = $ 2.94 orders of magnitude. Thus, there appears be a slight degradation of the sample's conductivity range. A film-thickness measurement was taken before processing with a 632.8\,nm ellipsometer and found to be $\textrm{T}_{\textrm{VO2}} = $ 18.26\,nm using a measured index of refraction published in \cite{opt}. A measurement was taken post-processing with an atomic force microscope and found to have an average thickness of 17\,nm. The final material values used in the electromagnetic model that best fit the measured s-parameters are $\textrm{T}_{\textrm{VO2}} = $ 17\,nm, $\sigma_{\textrm{ON}} = $ 10$^5$\,S/m and $\sigma_{\textrm{OFF}} = $ 80\,S/m. The conductivity of gold was also measured at $\sim$25$^\circ$C and found to be $\sigma_{\textrm{Au}} = $ 4.87$\times$10$^7$\,S/m.

\section{Measurements and Unit-cell Modeling} \label{sec:measurement}

\subsection{Calibration and De-embedding}

Once the fabrication was complete, s-parameter measurements were taken well above and below the {{\VO}} transition temperature of $\sim$68$^\circ$C (hot: $\sim$85$^\circ$C, cold: $\sim$25$^\circ$C) from 10\,MHz to 50\,GHz with an IF bandwidth of 100\,Hz using a Keysight PNA. The sample temperature was measured by a Fluke IR meter and heated well beyond the temperature at which s-parameters stabilized. The sample was placed on a carbon-impregnated AlN absorber (Fig.\,\ref{fig:sample}b) to eliminate any possible ground effects due to the metal chuck of the probe station. %Care was taken to land and skate the probes equal distances on all structures measured. 

While a multi-line calibration kit (e.g,. multi-line TRL) would enable s-parameter measurements calibrated to the edge of the 8 unit-cells, limited space on the {\VO} sample chip prevented patterning more than a single THRU standard. Therefore, the measurement setup's systematic errors were calibrated out up to the probe tips using LRRM standards on an impedance standard substrate (ISS) and launches were de-embedded with an assumption of symmetry. However, the ISS has different pad parasitics than the fabricated sample: the ISS has different fringe (shunt) capacitance due to different dielectric constant and waveguide dimensions, and different series inductance and contact resistance due to thicker metalization and different waveguide dimensions. While it is not straightforward to de-embed all the remaining pad parasitics, de-embedding the contact resistance is simple. At DC, a transmission line is simply a series resistor R, whose ABCD matrix directly gives this resistance and is of the following form:

\begin{equation}
\begin{bmatrix}
A & B\\
C & D\\
\end{bmatrix}
=
\begin{bmatrix}
1 & R\\
0 & 1\\
\end{bmatrix}
\end{equation}

At frequencies slightly higher than DC, the resistance of a short ($\ll$ $\lambda$) transmission line will increase slightly, becoming complex, yet will remain an excellent representation of the line. Therefore, the THRU standard was modeled electromagnetically at 10 MHz, the series R was extracted, and the contact resistance was de-embedded. This contact resistance was assumed to be the same for all structures.

To extract the unit-cell response, the whole structure was treated as a linear system of cascaded individual, symmetric structures represented by transmission parameters. In this paper, T represents ABCD parameters. The unit-cell was then solved for using matrix inversion and matrix square root operations. 

After removing contact resistance, the launches and (remaining pad parasitics) were de-embedded using  
\begin{equation}\label{eq:launch}
    \text{\scalebox{1}{$[\textrm{T}_{\textrm{Launch}}] = [\textrm{T}_{\textrm{Thru}}]^{1/2}$}}\\
\end{equation}
by taking the matrix square root of the entire THRU structure and assigning identical left and right launches to the post-processing of all consequent structures:
\begin{comment}
\begin{equation} \label{eq:linearstructure}
    \text{\scalebox{1}{$[\textrm{T}_{\textrm{struct}}] = [\textrm{T}_{\textrm{Launch}}][\textrm{T}_{\textrm{UC}}]^{8}[\textrm{T}_{\textrm{Launch}}]$}}\
\end{equation}
\end{comment}
\begin{equation}\label{eq:deembed}
    \text{\scalebox{1}{$[\textrm{T}_{\textrm{UCx8}}]=[\textrm{T}_{\textrm{Launch}}]^{-1}[\textrm{T}_{\textrm{struct}}] [\textrm{T}_{\textrm{Launch}}]^{-1}$}}\
\end{equation}

\begin{figure}[t]
\centering
\includegraphics[width=\columnwidth]{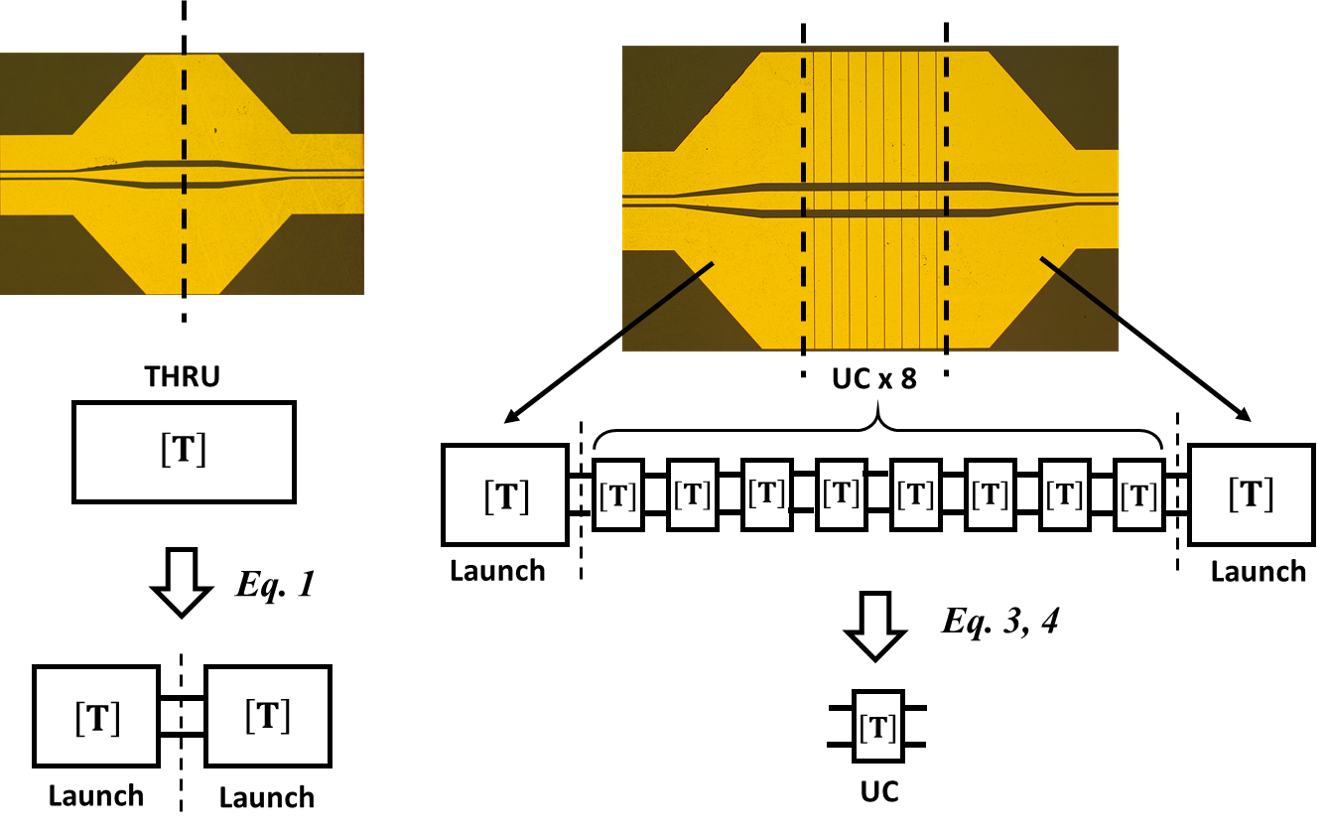}
\caption{Diagram illustrating de-embedding and unit-cell extraction from (\ref{eq:launch}), (\ref{eq:deembed}), and (\ref{eq:UCextraction}). All blocks are represented by their respective transmission matrix \textbf{T}. }
\label{fig:DeEmbed}
%\vspace{-0.5cm}
\end{figure}

This assumes that the left and right probe landings were exactly repeatable for all structures. This procedure is summarized in Fig.\,\ref{fig:DeEmbed}. Repeatability is estimated to be within $\pm$\,15\,$\mu$m, which corresponds to a phase-shift of $\pm$\,2.1$^\circ$ at $50\,$GHz. Additionally, de-embedding launches with pad parasitics taken from a single THRU measurement produces artifacts in the data, as will be explained in detail in section \ref{subsec:results}.

To obtain individual unit-cell parameters, the remaining 8 unit-cell structure was processed using (\ref{eq:UCextraction}). Since taking the matrix square root is equivalent to obtaining the parameters for half the original length, taking the matrix square root 3 times gives parameters for $\frac{1}{2^3} = \frac{1}{8}$ length or one unit-cell; this is the advantage of measuring lengths that are powers of 2.
\[\text{\scalebox{1}{$[\textrm{T}_{\textrm{UCx4}}] = [\textrm{T}_{\textrm{UCx8}}]^{1/2}$}}\]
\[\text{\scalebox{1}{$[\textrm{T}_{\textrm{UCx2}}] = [\textrm{T}_{\textrm{UCx4}}]^{1/2}$}}\]
\begin{equation}\label{eq:UCextraction}
    \text{\scalebox{1}{$[\textrm{T}_{\textrm{UC}}] = [\textrm{T}_{\textrm{UCx2}}]^{1/2}$}}\
\end{equation}
\noindent The assumption here is that there are no differences between unit-cells; in reality, any existing differences due to fabrication are averaged across all unit-cells.

\subsection{Results and Modeling} \label{subsec:results}

\begin{figure*}
    \includegraphics[width=\linewidth]{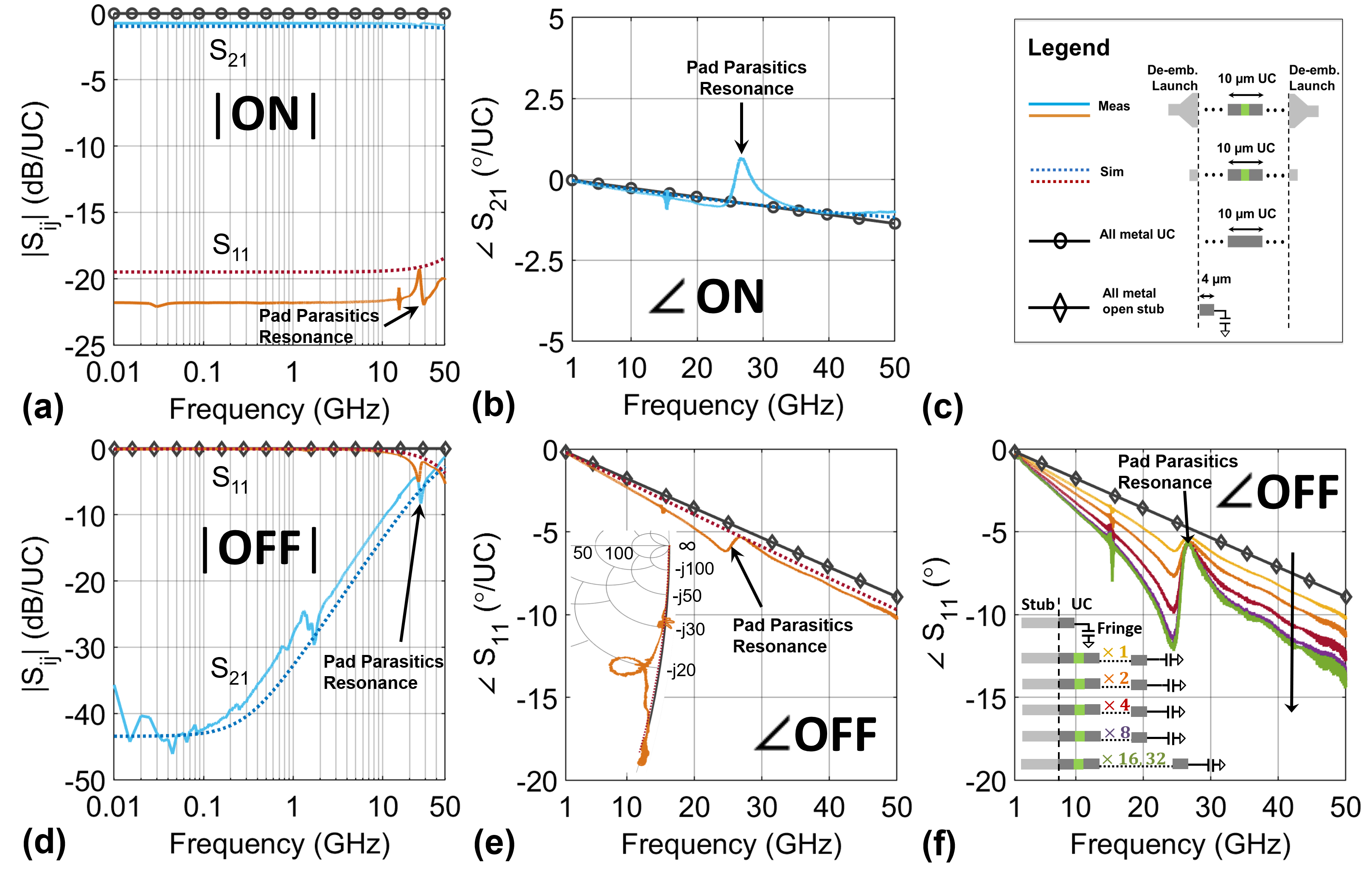}     
    \caption{S-parameters for structure 2 (Table\,\ref{tab:designparam}): solid is measured, dotted is simulated using MoM. Top row (a, b, c) and bottom row (d, e, f) show ON-state and OFF-state results, respectively. Black curves with circular markers for (a) and (b) correspond to $S_{21}$ of a 10 $\mu$m pure metallic line. Black curves with diamond markers for (d) and (e) correspond to $S_{11}$ of a 4 $\mu$m metallic open stub with fringe capacitance. (c) serves as the legend for the entire figure. (f) shows the phase of the reflection coefficient converging after 16x OFF-state UCs + simulated fringe capacitance are cascaded at the end of a 4 $\mu$m stub. Plots (e) and (f) show the 2-port measured OFF-state UCs converted to 1-port measurements, with the second port terminated in simulated fringe capacitance.}
    \label{fig:struct2}
    \vspace*{-1pt}
\end{figure*}

\begin{figure}[ht!]
    \begin{tabular}{@{}l@{}l@{}}
    \includegraphics[width=\columnwidth]{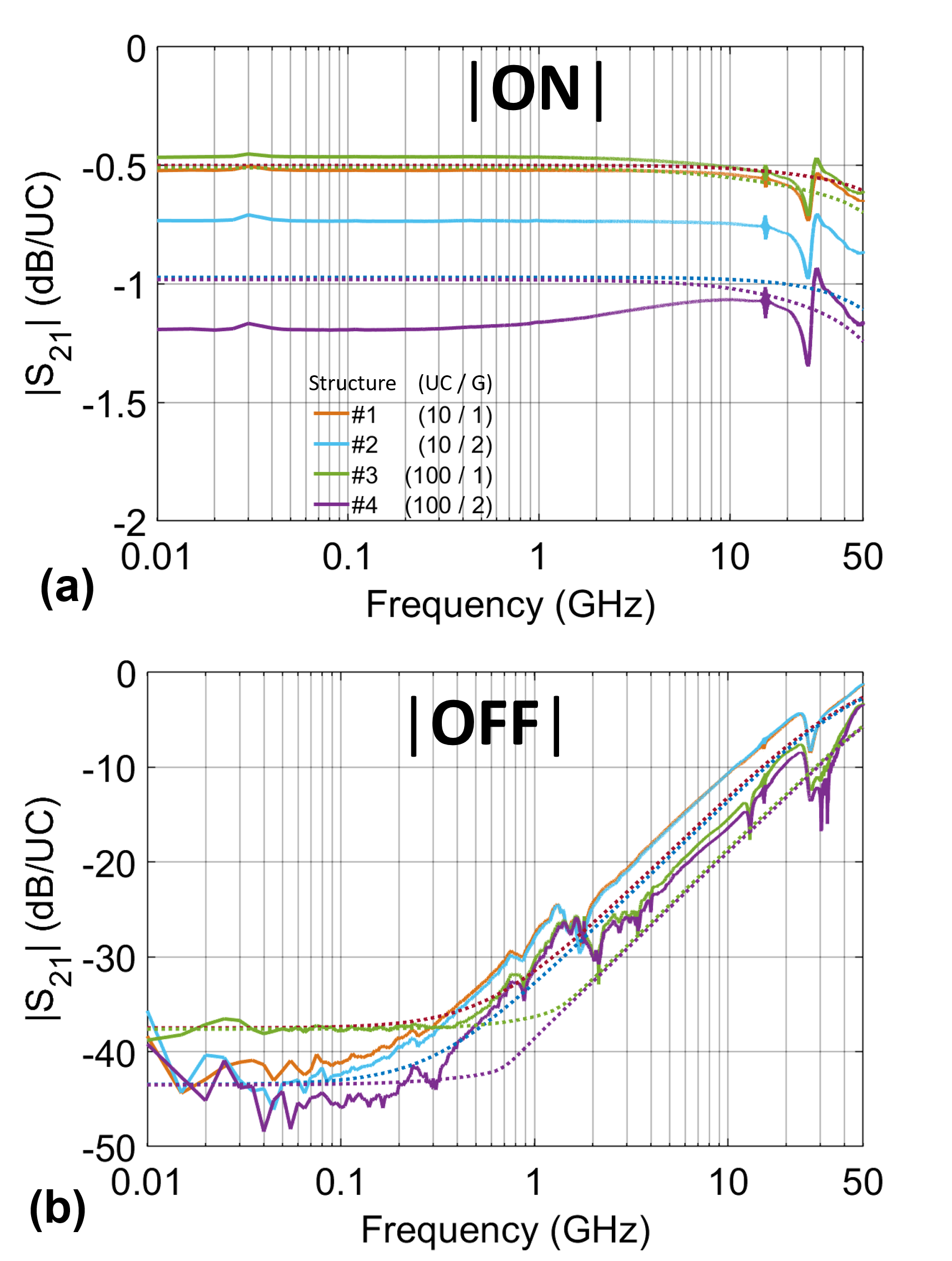}\
    \end{tabular}
    \caption{S-parameters for all fabricated structures (Table\,\ref{tab:designparam}): solid is measured, dotted is simulated using MoM. (a) shows the ON-state results, and (b) shows the OFF-state results.}
    \label{fig:structall}    
    \vspace{+4pt}
\end{figure}

Using the above approach, the unit-cell s-parameters of all structures were de-embedded. In order to understand the unit-cell response in detail, we first focus on a single structure, \#2 (Table \ref{tab:designparam}) with a 2\,$\mu$m gap and 10\,$\mu$m unit-cell length, and then discuss the other structures. Figure\,\ref{fig:struct2} shows the measured (solid) and simulated (dotted) s-parameters of the de-embedded unit-cell for structure \#2 in metallic and insulating states (labelled as ``ON'' and ``OFF'', respectively). A comparison of all structures may be seen in Fig.\,\ref{fig:structall}. Both electromagnetic models (Method of Moments) and circuit models were obtained using industry-standard simulation tools. The electromagnetic model was based on the measured {\VO} thickness and conductivities (section \ref{sec:fabrication}) and adjusted slightly to obtain good fits of the measured s-parameters as specified in Sec.\,\ref{sec:fabrication} above and repeated here for convenience. The final material values used in the electromagnetic model that best fit the measured s-parameters are $\textrm{T}_{\textrm{VO2}} = $ 17\,nm, $\sigma_{\textrm{ON}} = $10$^5$\,S/m and $\sigma_{\textrm{OFF}} = $ 80\,S/m. The conductivity of gold was also measured at $\sim$25\,$^\circ$C and found to be $\sigma_{\textrm{Au}} = $ 4.87$\times$10$^7$\,S/m. These are all included in the EM model in Fig.\,\ref{fig:UnitcellParameters}. The actual fabricated dimensions of the waveguide were also accounted for: S = 106.5\,$\mu$m and W = 45\,$\mu$m. The circuit model of Table\,\ref{tab:circuitparam} was fit to the measured s-parameters across the entire band, and the resulting values are summarized in Table\,\ref{tab:circuitparam}. The ON and OFF-state resistance of the {\VO} is modeled as a series resistance R\textsubscript{VO2}, and the gap capacitance is modeled by a series capacitance C\textsubscript{GAP} and shunt capacitances C\textsubscript{sh} representing fringing fields from the CPW signal line to ground planes \cite{naghed_equivalent,gevorgian_cad}. In addition to the {\VO} unit-cell, Fig.\,\ref{fig:struct2} shows the s-parameters of a simulated, realistic 10 $\mu$m pure metal line (black solid, circle marker)---the ideal response to which the ON-state {\VO} with inclusions should be compared---and a simulated, realistic 4\srev{ }\rev{\,}$\mu$m metal open stub with fringe capacitance (black solid, diamond marker)---the ideal response to which the OFF-state {\VO} with inclusions should be compared. This comparison indicates how well this reconfigurable concept compares to a traditional, non-reconfigurable circuit.

In general, the measured and simulated values are in good agreement, indicating that the {\VO} film parameters used in the EM models as summarized in Fig.\,\ref{fig:UnitcellParameters} are valid and useful for new designs. The measured transmission, $S_{21}$, undergoes a change of approximately four orders of magnitude between the ON and OFF states; however, the ON-state insertion loss is as much as $0.87$\,dB/UC up to 50\,GHz. As will be discussed below, this loss is dominated by the resistance of the {\VO} film due to the limited film thickness, 17\,nm, used in these proof-of-concept structures. The OFF-state isolation is better than 40\,dB at low frequencies, but as frequency increases, the isolation follows a nearly perfect series R\,$||$\,C response \rev{(i.e., the isolation matches that of a parallel R\,$||$\,C network in series with the two ports of the VNA)}. These observations indicate that there is a fundamental trade-off between loss and isolation across the unit-cell gap but that increased film thickness can mitigate the loss, as will be discussed in section\,\ref{sec:implementation}. However, before further discussion of the measurements, we address the presence of three artifacts in the measurement.

There are three observable resonances (1.5, 15.35, and 26.4\,GHz) in the measured results, and each of these correspond to calibration and de-embedding limitations, not to the unit-cell structures. The resonance seen near 26.4\,GHz is due to differences in pad parasitics between the ISS structures and the test structures on sapphire which manifest when computing (\ref{eq:launch}). Due to limited space on the {\VO} sample, the THRU standard was limited to 2.5\,mm long, where it is $\frac{\lambda}{2}$ at 26.4\,GHz ($\epsilon_{\textrm{eff}} = 5.15$). A transmission line, at multiples of $\frac{\lambda}{2}$, does not transform the load impedance. Instead, the line will appear transparent, with exception of accumulated phase and loss. Since the THRU standard is terminated with pad parasitics and 50\,$\Omega$ (calibrated system) load, the input impedance at 26.4\,GHz will be simply the parasitics and calibrated system. Therefore, any method of de-embedding launches from a THRU that still has parasitics will cause a resonance to appear \cite{heymann_deembedding_1994,mangan_deembedding_2006,haydl_design_1993}. When modeling this entire structure and de-embedding using the same procedure as in Fig.\,\ref{fig:DeEmbed}, the same resonance appears at 26.4\,GHz. Adjusting the launch length of the simulated THRU standard to 0.25\,mm causes the resonance to be pushed to a much higher frequency outside the measurement range. Therefore, no resonance appears in the simulated curves in Fig.\,\ref{fig:struct2}.

The $S_{21}$ of the {\VO} unit-cell of structure 2 (Fig.\,\ref{fig:struct2}(a)) in the ON state is at most $\textrm{-0.867}$\,dB per unit-cell across the entire measured band, while the pure metal unit-cell is at most $\textrm{-0.0022}$\,dB per unit-cell. Obviously, this is impractical; this is the result of very thin {\VO} films (17\,nm) and a large unit-cell gap (2\,$\mu$m). As will be shown later, smaller gaps and thicker films exhibit losses closer to that of a pure metal line, thereby demonstrating potential for this reconfigurable concept to work.

The lines' phase progressions (Fig.\,\ref{fig:struct2}(b)) are more similar with $\beta l =$ 1.03$^\circ$ and 1.371$^\circ$ at 50\,GHz for a {\VO} and metal unit-cell, respectively. This deviation of the measured {\VO} unit-cell from the pure metallic unit-cell at 50\,GHz (and other frequencies) is partially due to distortion by the pad parasitics nearing the frequency (52.8\,GHz) at which the line is $\lambda$. The electromagnetic simulation shows a slight dispersion of the phase for a {\VO} UC, which makes more sense given the significant loss of the oxide.

Nonetheless, low ON-state insertion loss is not the only criterion necessary to render this concept practical: high OFF-state isolation is also essential to create well-defined current paths. The isolation of a {\VO} unit-cell (Fig.\,\ref{fig:struct2}(d)) in the OFF state is greater than 40\,dB below 200\,MHz and degrades with a slope of $\sim$20\,dB/dec as frequency increases, the same slope as a series capacitor. The high $S_{11}$ indicates the power is primarily reflected, not absorbed, across the entire band. Below 200\,MHz, the {\VO} film dominates the insertion loss due to its low conductivity, and above 200\,MHz, the unit-cell gap capacitance dominates. In fact, the OFF-state unit-cell is modeled extremely well by a series resistor R\textsubscript{VO2} in parallel with a capacitor C\textsubscript{GAP}, and the values are found in Table\,\ref{tab:circuitparam}. This means that higher frequencies are worse at confining current to a designed path and allow current to leak into nearby unit-cells. This is extremely relevant to designing antennas, where the current distribution dictates the far field radiation pattern. This is also pertinent to designing the characteristic impedance of distributed circuits, which is inherently dependent on widths of and gaps between conductors. The characteristic impedance will exhibit an increased frequency dependence, and this must be accounted in the design of these circuits.

\begin{table*}[t]
\renewcommand{\arraystretch}{1.3}
\caption{Unit-cell Transmission Line and Circuit Parameters.}
\label{tab:circuitparam}
\centering
\begin{tabular}{@{}cccc|ccc@{}}
    \toprule
    Structure (UC\,/\,G\,$\mu$m) & Z\textsubscript{o}\,($\Omega$) & $\alpha$\,(dB/mm) & $\beta$\,(rad/mm) & R\textsubscript{VO2}\,($\Omega$) & C\textsubscript{GAP}\,(fF) & C\textsubscript{sh}\,(fF)\\
     &    ON      & ON  & ON &    ON / OFF        &   ON\&OFF               & ON\&OFF \\
    \midrule
        \textbf{Metal (10\,/\,0)} &  \textbf{50.6 - j0.520} & \textbf{0.217} & \textbf{2.39} & \textbf{0.0252} &  & \\
   \vhtable     1 (10\,/\,1) & 101 - j80.5 & 31.3 & 4.47 & 5.92 / 7,288 & 35.6 & 0.868\\
   \vhtable     2 (10\,/\,2) & 128 - j119 & 46.4 & 5.69 & 11.8 / 14,566 & 34.2 & 1.56\\
    \vhtable    3 (100\,/\,1) & 60.4 - j16.9 & 5.98 & 2.57 & 5.91 / 7,465  &  17.9 & 0.220\\
    \vhtable    4 (100\,/\,2) & 64.4 - j28.8 & 10.3 & 2.75 & 11.8 / 15,035 &  17.3 & 0.450\\
    \midrule
        Structure (T\textsubscript{VO2}\,/\,G\,nm): $\textrm{UC}=10\,\mu$m & & & & & &\\
    \midrule
   \vhtable     100\,/\,100 & 56.2 - j6.26 & 2.52 & 2.51 & 0.103 / \srev{125.4}\rev{2,003} & 42.1 & 0.200\\
   \vhtable     100\,/\,500 & 59.9 - j15.6 & 6.15 & 2.67 & 0.506 / \srev{627}\rev{10,000} & \srev{38.3}\rev{37.8} & 0.402\\
    \vhtable    200\,/\,100 & 55.5 - j4.71 & 1.91 & 2.47 & 0.0522 / \srev{62.8}\rev{1,003}  &  \srev{41.5}\rev{43.3} & 0.157\\
    \vhtable    200\,/\,500 & 57.9 - j9.99 & 3.96 & 2.58 & 0.254 / \srev{314}\rev{5,010} &  \srev{38.7}\rev{38.2} & 0.256\\
    \bottomrule
%    \vspace*{+1pt}
    \vhtable       &  \multicolumn{3}{c}{\includegraphics[width=0.5\columnwidth]{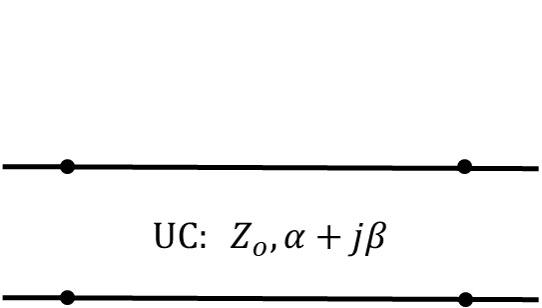}} & \multicolumn{3}{c}{\includegraphics[width=0.5\columnwidth]{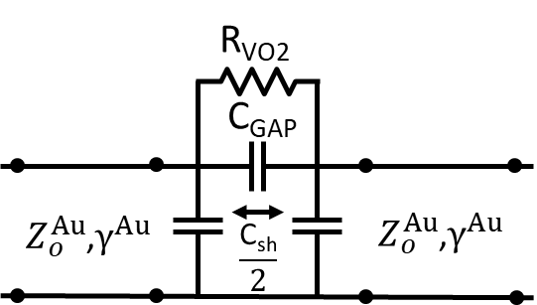}}\\
\end{tabular}
    \vspace*{-1pt}
\end{table*}

Because a {\VO} unit-cell in the OFF-state is supposed to approximate a metallic open stub with fringe capacitance, a comparison of the phase of the reflection coefficient is also very indicative of the OFF-state behavior. In order to compare with a 1-port metallic open stub, the 2-port OFF-state unit-cell was converted to a 1-port measurement. However, the original 2-port OFF-state unit-cell measurement did not include fringe capacitance, so the second port was terminated in a simulated fringe capacitance. In the case of an entire chip of embedded metallic inclusions in a PCM film, the OFF-state phase of the unit-cell is not only determined by the first OFF-state gap but also by all the consecutive unit-cells which will couple through fringing fields. Each unit-cell will provide isolation but also add slightly to the phase of $S_{11}$ because field lines will also terminate on conductors in distant unit-cells (albeit with diminishing proportion). Fig.\,\ref{fig:struct2}(f) illustrates this by plotting the reflection coefficient phase for 1 through 32 cascaded UCs in the OFF-state. The phase converges at 16 UCs; it is hidden beneath the 32 unit-cell curve. Therefore, the total additional phase accumulated by a stub terminated in these OFF-state UCs is dominated by the first 16 UCs and is $\Delta\phi = 5.06^\circ$ at 50\,GHz for this structure. If this extra phase were represented by an increase in fringe capacitance, then $\Delta\textrm{C} = 7.6-4.97 = 2.63$\,fF. This additional capacitance must be considered when designing metallic/PCM-based stubs, such as in matching networks, where the required stub length should be reduced slightly to account for the added capacitance.

All the measured structures are compared together in Fig.\,\ref{fig:structall}. Figure\,\ref{fig:structall}(a) shows two trends in the ON-state. First, measured (solid curves) structures 1 and 3 have very similar loss even though they are 10\,$\mu$m and 100\,$\mu$m long, respectively, because s-parameters are plotted in dB/UC and the gaps are identical. Second, measured structures 2 and 4 show more loss than 1 and 3 due to a larger gap (2 vs 1\,$\mu$m), but their loss does not double like intuition and simulation (dotted) would predict. This is attributed to non-uniformity in film thickness and/or quality across the sample. Figure\,\ref{fig:structall}(b) also shows two trends in the OFF-state. At low frequencies, structures 1 and 3 have less isolation than 2 and 4, although 1 and 3 should be nearly identical as indicated by the simulated curves. Again, this discrepancy is attributed to non-uniformity in the film. At high frequencies, structures 1 and 2 have similar (although not identical) capacitance, whereas structures 3 and 4 have similar capacitance. A change in gap G does not affect the capacitance much because, for this metal thickness and geometry, the capacitance is due mostly to fringing fields: fields not directly in the gap cross-section but beginning and terminating on the metallic inclusions. Structures with the same gap (1 and 3) differ significantly in capacitance C$_{\textrm{GAP}}$ confirming that metal in close proximity to the gap dominates the capacitance instead of the gap (for these gap and line geometries).

\rev{
\begin{figure}[t]
    \includegraphics[width=\columnwidth]{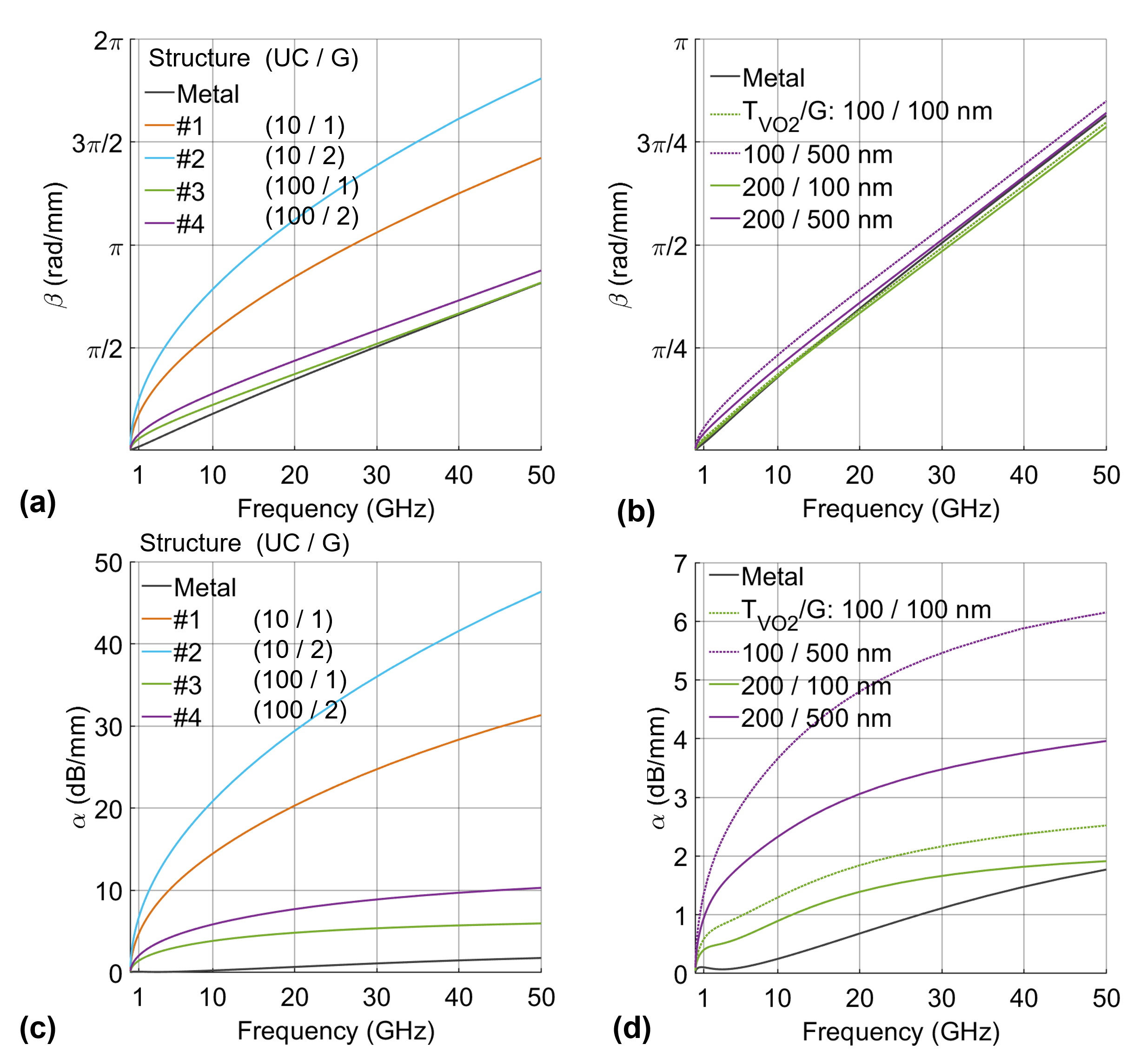}\\
    \caption{\rev{(a) and (c) Phase and attenuation constant per millimeter as a function of frequency for each of the fabricated 17\,nm {\VO} structures. (b) and (d) Phase and attenuation constant per millimeter as a function of frequency for a simulated unit-cell of thicker {\VO} films and smaller gaps.}}
\label{fig:dispersion}
    %\vspace{-0.5cm}
\end{figure}
}

\rev{A transmission line model (bottom left figure in Table\,\ref{tab:circuitparam}) is extracted from the EM simulation of all fabricated and measured ON-state unit-cells. The attenuation constant, $\alpha$, and propagation constant, $\beta$, are plotted versus frequency in Fig.\,\ref{fig:dispersion}(a) and (c). In general, the very thin {\VO} causes the attenuation constant, $\alpha$, to be very high compared to a metal-only unit-cell and to increases as the {\VO} percentage increases. There is significant dispersion evident in Fig.\,\ref{fig:dispersion}(a) for structures 1 and 2 (UC $= 10\,\mu$m), suggesting that the propagation mode is very different from a quasi-TEM mode; some dispersion is evident for structures 3 and 4 (UC = 100\,$\mu$m) but is less than for structures 1 and 2 because of a lower percentage of {\VO} per millimeter. A transmission line model is given at 50\,GHz in the top left half of Table\,\ref{tab:circuitparam} which is useful for design from low-bands to 50\,GHz for the 100\,$\mu$m unit-cell but it cannot be extrapolated to all lower frequencies for the 10\,$\mu$m unit-cell because of the dispersion. In general Fig.\,\ref{fig:dispersion} should be consulted for accurate modeling of transmission line propagation coefficients.}

\rev{Thicker films and smaller gaps (bottom left half of Table\,\ref{tab:circuitparam}), which are discussed more in section \ref{sec:implementation}, bring the attenuation constant to within three times that of metal (Fig.\,\ref{fig:dispersion}(d)), and therefore exhibit minimal dispersion in the phase constant, $\beta$ (Fig.\,\ref{fig:dispersion}(b)). The attenuation constant follows a square root dependency on frequency. Likewise, under the same conditions that bring $\alpha$ and $\beta$ closer to ideal, the characteristic impedance, $\textrm{Z}_o$, will remain similar to a metal unit-cell.}

\rev{A circuit model (bottom right figure in Table\,\ref{tab:circuitparam}) is fit to the EM simulation of all ON- and OFF-state unit-cells across the entire band. The metallic portion of each structure is modeled as a traditional transmission line ($\textrm{Z}_0^{\textrm{Au}}, \gamma^{\textrm{Au}}$) at each frequency point, while the {\VO} portion is modeled with a lumped resistor R\textsubscript{VO2} and a lumped gap capacitor model for the entire band. The gap capacitor model is assumed to be identical in both ON\,/\,OFF states; the C\textsubscript{sh} was fit in the ON-state and C\textsubscript{GAP} was fit in the OFF-state. For thick films, the circuit model is a very good fit, particularly the phase of S$_{21}$ for ON-state unit-cells, indicating that any parasitics present due to evanescent modes as a result of the discontinuities are either well represented or negligible. The {\VO} resistance scales linearly with gap, but the capacitance does not. Because the metal is so thin and the gap is so small, fringing fields dominate the capacitance and thus do not scale linearly with gap size. This is discussed more in section \ref{sec:implementation}. The C\textsubscript{sh} is much smaller than C\textsubscript{GAP} because the CPW gap W is much larger than the unit-cell gap G. Both these trends are predicted by a FEM model of a single unit-cell.}

\srev{A transmission line model (bottom left figure in Table\,\ref{tab:circuitparam}) is extracted at 50\,GHz from the EM simulation of all ON-state unit-cells (top left half of Table\,\ref{tab:circuitparam}). In general, due to the very thin {\VO}, the attenuation constant, $\alpha$, is very high compared to a metal-only unit-cell and increases as {\VO} percentage increases. Thicker films and smaller gaps (bottom left half of Table\,\ref{tab:circuitparam}), which are discussed more in section \ref{sec:implementation}, bring the loss significantly closer to metal. The phase constant, $\beta$, remains similar to metal for thick {\VO} films and high metal-oxide ratios unless the loss is extremely high, such as for structure 1 and 2. One can compensate for slight changes in $\beta$ by adjusting design lengths. Likewise, under the same conditions that bring $\alpha$ and $\beta$ closer to ideal, the characteristic impedance, $\textrm{Z}_o$, will remain similar to a metal unit-cell.}

\srev{A circuit model (bottom right figure in Table\,\ref{tab:circuitparam}) is fit to the EM simulation of all ON- and OFF-state unit-cells across the entire band. The metallic portion of each structure is modeled as a traditional transmission line ($\textrm{Z}_0^{\textrm{Au}}, \gamma^{\textrm{Au}}$) at each frequency point, while the {\VO} portion is modeled with a lumped resistor R\textsubscript{VO2} and a lumped gap capacitor model for the entire band. The gap capacitor model is assumed to be identical in both ON\,/\,OFF states; the C\textsubscript{sh} was fit in the ON-state and C\textsubscript{GAP} was fit in the OFF-state. The {\VO} resistance scales linearly with gap, but the capacitance does not. Because the metal is so thin and the gap is so small, fringing fields dominate the capacitance and thus do not scale linearly with gap size. The C\textsubscript{sh} is much smaller than C\textsubscript{GAP} because the CPW gap W is much larger than the unit-cell gap G. Both these trends are predicted by a FEM model of a single unit-cell.}

\begin{figure}[b]
    \centering
    \includegraphics[width=\columnwidth]{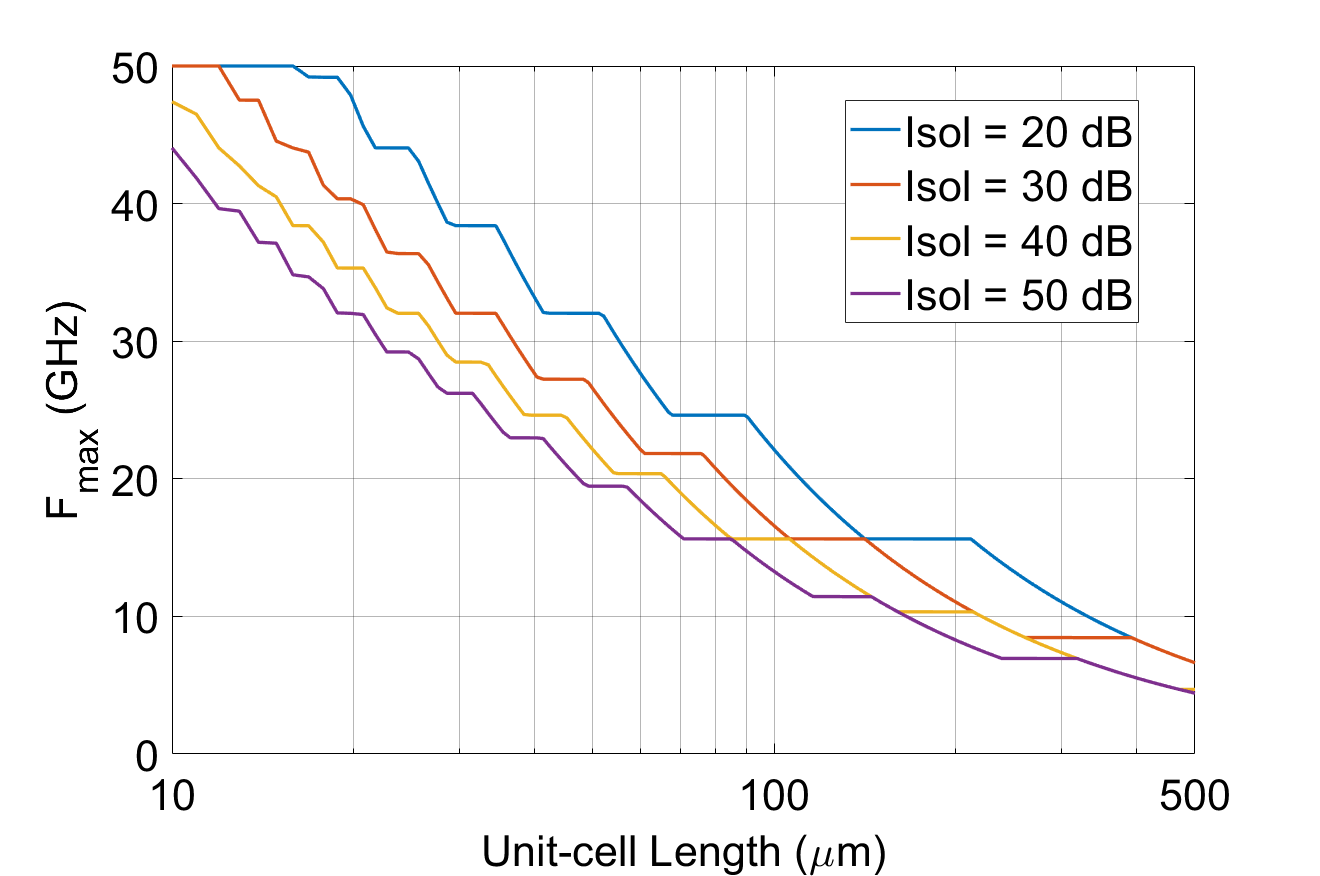}
    \caption{\rev{Maximum operating frequency versus unit-cell length based upon a requirement to achieve 20, 30, 40, and 50\,dB of isolation within an electrical length of $\lambda/20$. For the $10\,\mu$m ($100\,\mu$m) cells used in this work, $f_{max}=45\,$GHz ($15\,$GHz) for a 40\,dB isolation requirement. R\textsubscript{ON} and C\textsubscript{OFF} from the measured 17\,nm-thick {\VO} films were used in this calculation but the results are almost the same for thicker films since C\textsubscript{OFF} ends up dominating the high-frequency response.}}
    \label{fig:fmax}
\end{figure}

\rev{\subsection{Maximum Operating Frequency}} 

\rev{While ON-state insertion loss is relatively low and constant versus frequency up to 50\,GHz (see Fig.\,\ref{fig:structall}), this does not necessarily indicate these structures are useful up to 50\,GHz because OFF-state isolation, which is dominated by capacitance, reduces significantly as frequency increases. In this section we are interested in a rule for maximum operating frequency of the proposed structures. In general, a single OFF-state unit-cell does not need to provide the total desired isolation in order to realize a particular circuit geometry. Rather, unit-cells may be cascaded together to achieve a high degree of total isolation if the unit-cells are small compared to a wavelength. For example, consider that a $10\,\mu$m ($100\,\mu$m) unit-cell is $\lambda/20$ at 660\,GHz (66\,GHz); therefore, many cells can participate in accomplishing the required isolation for frequencies below these values. Consequently, we propose a criterion to determine the maximum operating frequency of the proposed structures based upon the length required to achieve a desired isolation in the OFF-state. Specifically, we define the maximum frequency of operation to be that frequency at which the desired isolation is achieved within $\lambda/20$. This means that the correct number of unit-cells and the correct unit-cell length must be selected to achieve the required isolation within $\lambda/20$.}

\rev{In order to determine the maximum operating frequency $f_{max}$ versus unit-cell length, we first compute the number of unit-cells required to achieve a particular isolation specification (here 20, 30, 40, 50\,dB) at a given frequency and then compute the required unit-cell length so that the overall length is less than $\lambda/20$ at the chosen frequency. Figure\,\ref{fig:fmax} shows the resulting $f_{max}$ versus unit-cell length using the R\textsubscript{ON} and C\textsubscript{OFF} values extracted from measurements of the 17\,nm-thick {\VO} films in this work. However, we point out that these curves are very similar for thicker (e.g., 200\,nm) films because the high-frequency limit is determined by C\textsubscript{OFF}. To satisfy a $40\,$dB isolation requirement, a $10\mu$m unit-cell length can operate up to $45\,$GHz and a $100\mu$m unit-cell length can operate up to $14\,$GHz. Note that the curves are discontinuous at some frequencies because $\lambda/20$ has to be an integer multiple of a unit-cell length in order to meet (or exceed) the isolation requirement. It is worth pointing out that some applications may not require such a stringent isolation spec or they may be robust to vestigial lines longer than $\lambda/20$, in which case the maximum frequency would increase.}

\section{Practical Implementation} \label{sec:implementation}

\subsection{Improved Electrical Performance}

\begin{figure}[ht!]
    \centering$
    \begin{tabular}{@{}c@{}}
    \includegraphics[width=\columnwidth]{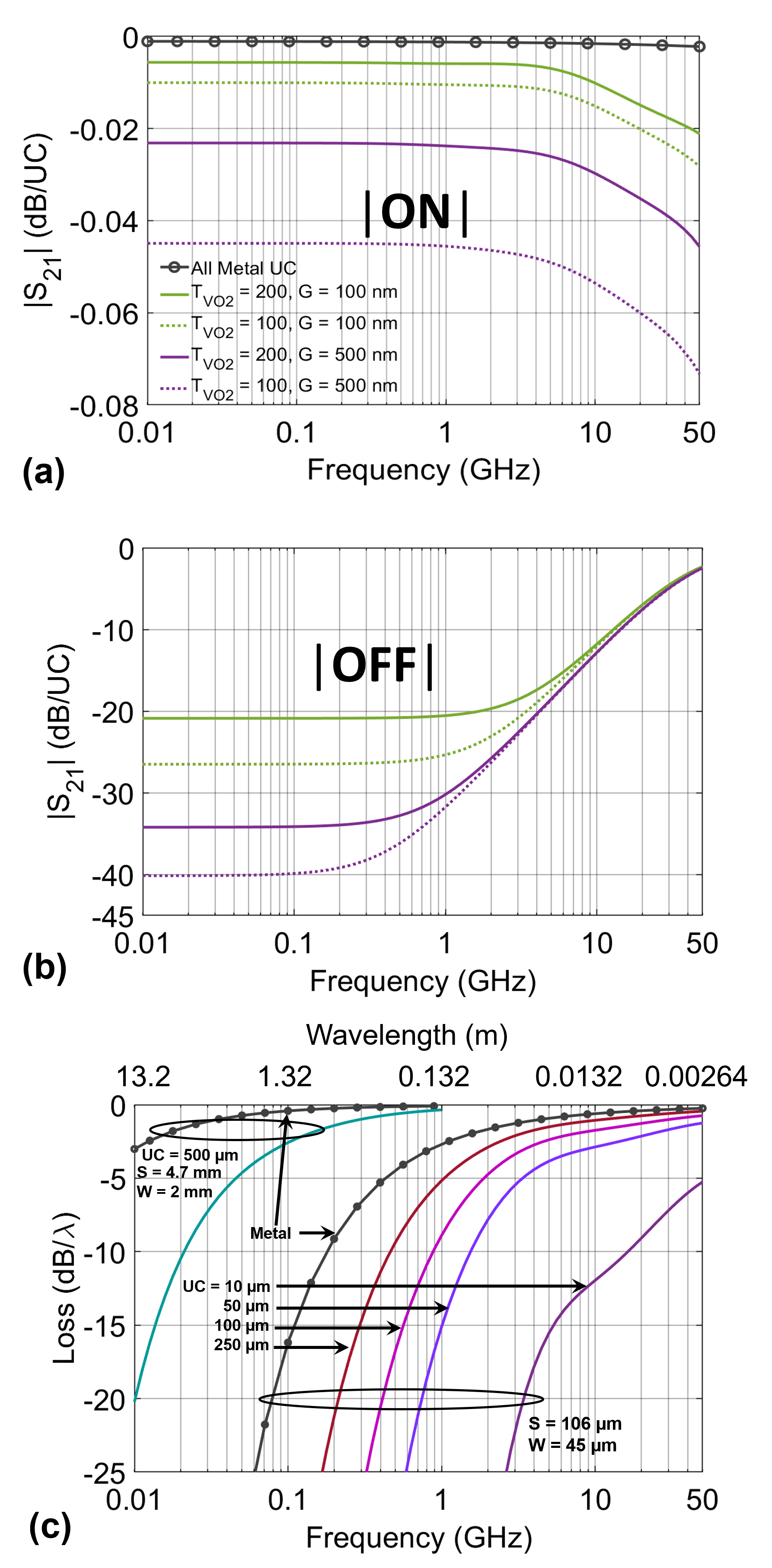}\\
    \end{tabular}$
    \caption{\rev{In order to feasibly design practical low-loss circuits, parametric studies were conducted of T\textsubscript{VO2} and G for ON state (a) and OFF state (b). In (c), curves shows the $|S_{21}|$ loss for CPW lines of length equal to the corresponding wavelength at each frequency, where the lines comprise unit-cell sizes of 10, 50, 100, or 250\,$\mu$m. Curves shown are for T\textsubscript{VO2} = 200\,nm and G = 100\,nm. Metal lines (black, cicle marker) are shown for reference. Structures with signal width S = 106.5\,um and gap G = 45\,um are inherently suited for microwave and millimeter-wave lengths, whereas larger CPW dimensions of S = 4.7\,mm and G = 2\,mm are suited for longer wavelengths. All curves are with T\textsubscript{Au} = 2\,$\mu$m.}}
    \label{fig:ParamLambda}
\end{figure}

The previous results \rev{highlight}\srev{yield two important conclusions, i)} the need for thicker films and smaller gaps to make this concept feasible\srev{, and ii) a validation of our ability to electromagnetically model these films and gaps with a high degree of confidence}. Consequently, two additional 10\,$\mu$m unit-cell structures with realizable gap sizes and reasonable film thicknesses are simulated and presented in Fig.\,\ref{fig:ParamLambda}(a) and (b): G = 100 and 500\,nm, and T$_{\textrm{VO2}} =$ 100 and 200\,nm. \rev{Moreover, in these simulations the OFF-state conductivity of {\VO} has been lowered to 5\,S/m, which has been the measured conductivity for films above the critical thickness ($\sim$20\,nm) \cite{comb}.}

\rev{As expected, thicker films and smaller gaps dramatically reduce the loss. A film thickness of 200\,nm and gap of 100\,nm yield at most 0.012\,dB/UC insertion loss from 0.01--5\,GHz and a maximum loss of 0.028\,dB/UC at 50\,GHz, much closer to the metal's 0.0022\,dB/UC, in contrast to 0.867\,dB for structure 2. The price paid for this in the OFF-state is reduced isolation: 20\,dB/UC from 0.01--1\,GHz, 11.8\,dB/UC at 10\,GHz, and 2.3\,dB/UC at 50\,GHz. The metal thickness for all structures plotted in Fig.\,\ref{fig:ParamLambda}(a) and (b) was maintained at 200\,nm, but the metal thickness can be increased up to 2\,$\mu$m without significantly increasing the capacitive coupling between unit-cells. Additional FEM simulations show that, for gap sizes of 0.5 and 1\,$\mu$m, capacitance is nearly unchanging for a metal thickness up to 2\,$\mu$m and increases by 10$\%$ for a metal thickness of 5\,$\mu$m. For a gap size of 0.1\,$\mu$m, the capacitance increases by 23$\%$ for a metal thickness of 2\,$\mu$m and by 48$\%$ for a metal thickness of 5\,$\mu$m. The parallel-plate capacitance in the gap is quite small compared to the overall gap capacitance because the fringing fields, not the parallel-plate fields, dominate. When the metal thickness becomes large relative to the gap size (5\,$\mu$m metal thickness and 0.1\,$\mu$m gap), then the parallel-plate fields begin dominating.}
\srev{As expected, thicker films and smaller gaps dramatically reduce the loss. A film thickness of 100\,nm and gap of 100\,nm yields at most 0.012\,dB/UC insertion loss from 0.01--5\,GHz and a maximum loss of 0.028\,dB/UC at 50\,GHz, much closer to the metal's 0.0022\,dB/UC, in contrast to 0.867\,dB for structure 2. The price paid for this in the OFF-state is significantly reduced isolation: -6.69\,dB/UC from 0.01--10\,GHz and -3.15\,dB/UC at 50\,GHz.}

Because distributed circuits are functional at lengths comparable to a wavelength, it is most useful to see the loss incurred per the guided wavelength at each frequency. This is plotted in Fig. \ref{fig:ParamLambda}(c) for \rev{a} film thickness of \srev{100\,nm}\rev{200\,nm} and \rev{a} gap G of 100\,nm for several unit-cell sizes (10-purple, 50-violet, 100-magenta, \rev{250-dark-red,} and 500-cyan\,$\mu$m). A pure metallic line is shown (black, circle marker) as a reference. \rev{Metal thickness for all structures plotted in Fig.\,\ref{fig:ParamLambda}(c) were increased to 2\,$\mu$m, which is typical for mmICs, to show more realistic losses}. In general, the loss per wavelength decreases with increasing frequency because wavelength decreases inversely with frequency, whereas conductor loss increases by the square root of frequency. At 10\,GHz, a 10\,$\mu$m unit-cell has \srev{20}\rev{12} dB/$\lambda$ loss, whereas the \srev{50 and 100}\rev{50, 100, and 250}\,$\mu$m UCs have \srev{5.65 and 3.86}\rev{2.9, 1.74, and 1.06} dB/$\lambda$ loss, much closer to the metal's \srev{2.06}\rev{0.6} dB/$\lambda$. Thus, the unit-cell length is a very useful degree of freedom so long as the \srev{reconfigurable application allows for reduced resolution}\rev{overall isolation specification is met and the application allows for reduced spatial resolution.}

Up to now, the CPW geometry used in the measurements and simulations has been quite small in order to maintain a quasi-TEM mode at 50\,GHz, but such a small size has prohibitive insertion loss at low frequencies. Thus, a much larger geometry (S = 4.7\,mm, W = 2\,mm) was simulated \srev{up to}\rev{to target the band below} 1\,GHz with a 500\,$\mu$m unit-cell and plotted (aqua) alongside its metallic reference (black, circular marker). This geometry maintains losses less than 5 dB/$\lambda$ even down to \srev{140}\rev{44}\,MHz. \srev{All these numbers are conservative since the metal thickness for all structures plotted in Fig.\,\ref{fig:ParamLambda}a, b, and c has been maintained at 200\,nm. Thicknesses comparable to the skin depth would reduce losses even further.}

\subsection{Thermal Control Considerations}

\rev{While the primary purpose of this work is to explore the electrical performance of the PCM-metal surface-inclusion concept across the microwave and low-millimeter-wave band, we also provide a preliminary, quantitative study of a thermal control system similar to the heater array depicted in Fig.\,\ref{fig:Patchconcept}(a), in order to demonstrate the feasibility of such a system. Thermal control by spatial confinement of heat is challenging in thermally conductive substrates such as sapphire, but the use of a highly abrupt insulator-metal phase transition renders this approach feasible. Single crystal {\VO} films, such as the sample utilized in this paper, have a thermal transition width of 4-6 $^\circ$C per about 4 decades of conductivity \cite{comb,transition_vo2}. Beyond the 4 decades of nearly abrupt change the conductivity saturates quickly in the ON state, and in the OFF state the conductivity continues to lower very gradually with decreasing temperature (drops from 16\,S/m at 65\,$^\circ$C to 5\,S/m at 25\,$^\circ$C). Thus, we do not expect fluctuations in substrate temperature $\geq$70\,$^\circ$C or $\leq$66\,$^\circ$C to significantly affect the reconfigurable circuit parameters.}
    
    \begin{figure}[t]
    \centering
    \includegraphics[width=0.99\columnwidth]{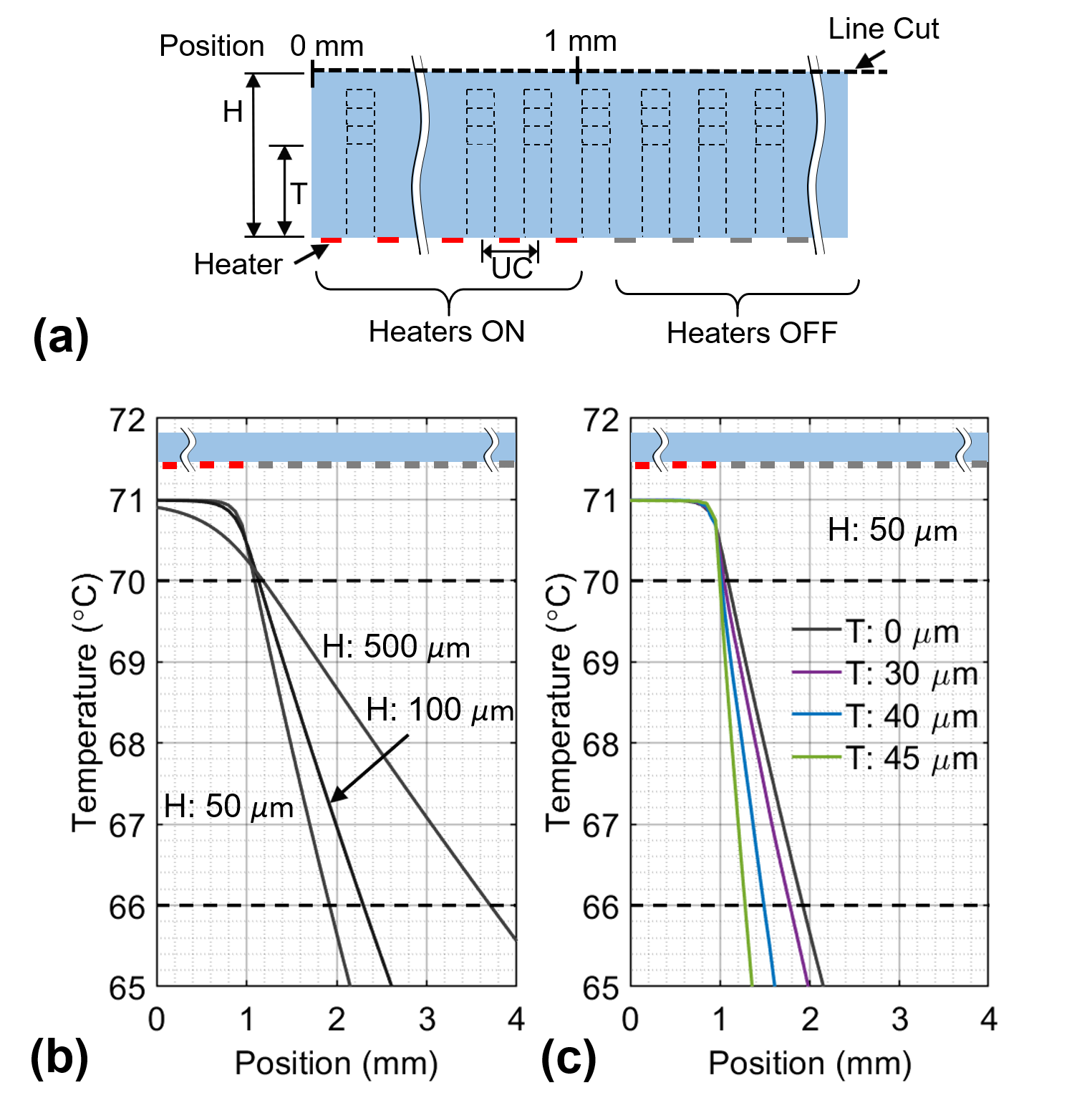}
    \caption{\rev{a) Thermal simulation setup with various sapphire substrate thicknesses (H), and various trench depths (T). Heaters at 71\,$^\circ$C are placed along the bottom of the substrate up to the 1\,mm mark. b) and c) plot thermal profiles vs. distance along top of substrate for various substrate thicknesses and trench depths.}}
    \label{fig:thermalSim}
    %\vspace{-0.5cm}
    \end{figure}
    
    \rev{Nonetheless, because the transition region is not instantaneous versus temperature, we have thermally simulated a 1D array of heat sources using an FEM heat-flow simulator (Ansys IcePak) to determine the heat profile in the substrate and the number of unit-cells needed to span the {\VO} transition region. Figure \ref{fig:thermalSim} shows the simulation setup and results. Figure\,\ref{fig:thermalSim}(a) shows the simulated geometry where control heaters are patterned on the back-side of a sapphire substrate (thickness H: 50,\,100,\,500\,$\mu$m) and spaced 100\,$\mu$m apart. Temperature was monitored along a line at the top of the substrate, where the unit-cell {\VO} would be present. Because {\VO} is nearly fully on above 70\,$^\circ$C, the heaters were specified as constant temperature regions at 71\,$^\circ$C. The heaters in Fig.\,\ref{fig:thermalSim}(a), (b), and (c) are colored red up to 1\,mm to indicate a temperature of 71\,$^\circ$C and are grayed out after 1mm because they do not have a temperature condition specified. Horizontal black dashed lines indicate the transition region (66-70\,$^\circ$C) of {\VO} to go from ON to OFF.}
    
    \rev{In order to improve confinement of heat within a unit-cell trenches of various depths were also modeled in the substrate (steep, straight sidewalls are easily realizable with DRIE as in our prior work \cite{Garcia_DRIE_2017}). In Fig. \ref{fig:thermalSim}(c), these trench depths are denoted as T in the legend. When T = 0 (black curves), no trench was etched. In general, as the substrate is thicker, more distance is required within the transition temperature region. By adding trenches, this adds isolation between each heater and reduces further the required distance to achieve a temperature transition. A trench distance of 90$\%$ of H (here, 45\,$\mu$m for H $=50\,\mu$m) is plotted as the best one could achieve, even though this is an impractical percentage to fabricate without compromising the substrate's mechanical integrity. A variety of more reasonable percentages are also shown ranging from 60\% to 90\% and the distances summarized in Table \ref{tab:thermalSum}.}
    
    \begin{table}[t]
\renewcommand{\arraystretch}{0.8}
\setlength{\tabcolsep}{4pt}
\caption{Thermal Simulations of Heater Arrays}
\label{tab:thermalSum}   
\centering
\begin{tabular}{rcccc}
    \toprule
    Substrate ($\mu$m) & 50 & 100 & 500 & \\
    \midrule
    \vhtable    Distance for T: 0\,$\mu$m & 0.83\,mm & 1.16\,mm & 2.53\,mm & \\
    \vspace*{+3pt}
    \vhtable    Frequency ($\lambda/20$) & 8.00\,GHz & 5.73\,GHz & 2.63\,GHz & \\
    \bottomrule
    \toprule
    Substrate 50\,$\mu$m, T ($\mu$m) & 0 & 30 & 40 & 45\\
    \bottomrule
    \vhtable    Distance & 0.83\,mm & 0.73\,mm & 0.48\,mm & 0.28\,mm\\
    \vspace*{+3pt}
    \vhtable    Frequency ($\lambda/20$) & 8.00\,GHz & 9.1\,GHz & 13.84\,GHz & 23.72\,GHz\\
    \bottomrule
%    \vspace*{+1pt}
\end{tabular}
    \vspace*{-1pt}
\end{table}

    \rev{The transition distance between fully ON and OFF unit-cells adds an effective length and resistance to the line. As a worst case, if we consider these unit-cells in the transition region to be fully on (maximum conductivity), this added length will only begin to matter at the frequency at which the length is $\lambda/20$ (according to the $f_{max}$ criterion we defined above). These frequencies are calculated and given in Table \ref{tab:thermalSum} for each variation in substrate size and trench depth. Specifically, for a 50\,$\mu$m substrate thickness, the distance required to transition to fully OFF (low conductivity) is 0.83, 0.73, 0.48, and 0.28\,mm, respectively. The corresponding frequency at which these lengths correspond to $\lambda/20$ are 8, 9.1, 13.8, and 23.7\,GHz, respectively. In reality, not all these unit-cells are on, but their resistance grows increasingly higher. This adds an additional isolation along the line so that the maximum operating frequency would be higher than that quoted in Tab.\,\ref{tab:thermalSum} but, in the case of a stub, this additional series resistance would result in a reduction of realizable $|\Gamma|$. 
    }
    
    \rev{Another consideration for a thermal control system is the significant temperature variation across a substrate due to, for example, hot-spots caused by power dissipation in active components. If such a hot-spot temperature is greater than the critical temperature of 68$^\circ$C and very close to a {\VO} unit-cell, then the cell may be unintentionally activated. Based on the same thermal simulations on sapphire discussed above, a heat source at 71\,$^\circ$C, which is $\Delta$T = 46\,$^\circ$C above ambient, required 100\,mm of distance for the substrate to reach ambient temperature (25\,$^\circ$C). This result can be used to estimate that 100\,mm is also required between an active component with a junction temperature of 117\,$^\circ$C (which is $\Delta$T = 46\,$^\circ$C above 71\,$^\circ$C), and the first OFF-state unit-cell after a heater. In reality, the distance required will be less than this, because the active component's $\Delta$T from ambient (92\,$^\circ$C) will increase its heat dissipation rate and cause the temperature to decay faster. Furthermore, in typical scenarios for the proposed {\VO} circuits, the {\VO} would be on a separate passive-only substrate from active devices (e.g., an impedance tuner wire-bonded to a PA). Therefore the sapphire-to-PCB transition would present a thermal discontinuity which would provide significant insulation, further mitigating this issue. Therefore, so long as {\VO} unit-cells are not placed in close proximity to heat sources, this problem can be mitigated.}
    
    \rev{Fabrication of a heater array should not require expensive or novel clean-room processes since it can be realized with standard lithographic processes and metal depositions (electron-beam evaporation or sputtering). A low-cost solution would be to fabricate the heater array on a separate, thin wafer, since this will eliminate the need for back-side wafer processing of the sample with the {\VO} film. An estimated power consumption from a thermal simulation of heaters on a 50\,$\mu$m sapphire reveals that 1.6\,$\mu$W was required to heat an area of 0.4\,mm$^2$ to 71$^\circ$. This is similar to thermal simulations in \cite{Gerislioglu_VO2AntennaWithMicroheaterArray_2017} where 1.5\,$\mu$W was sufficient to heat 4\,mm$^2$ to 87$^\circ$. Therefore, an estimated 30\,$\mu$W would be necessary to heat the reconfigurable dipole and 36\,$\mu$W would be necessary for the tunable triple-stub matching network presented in section \ref{sec:application}.}
    %(0.008\,mm$^2$,  (200 $\mu$W/mm$^2$))
    
\rev{\section{Applications}\label{sec:application}}

\rev{As seen in Fig.\,\ref{fig:fmax}, depending on the unit-cell size the proposed structures could perform well in the 10's of GHz and even up to 45\,GHz which opens up a broad range of applications. These include both antenna applications and guided-wave circuits. Just to list a few examples: reflectarrays \cite{Reflectarray} are often deployed on small satellites because of their compact size and high aperture efficiency. The proposed structures could be used to realize a dynamically reconfigurable version of these surfaces. Similarly, lens antennas are often used for satellite communications \cite{GRINLens} due to their high gain and wideband operation. The proposed structures could be used in these switch-beam lens antenna systems to greatly simplify the switch matrix used as a feed network. And of course dynamic matching networks such a stub-tuners \cite{Rebeiz_Tuner} could provide for high efficiency operation of power amplifiers under varying load impedance conditions.}

\rev{
    \begin{figure}[t]
    \centering
    \includegraphics[width=\columnwidth]{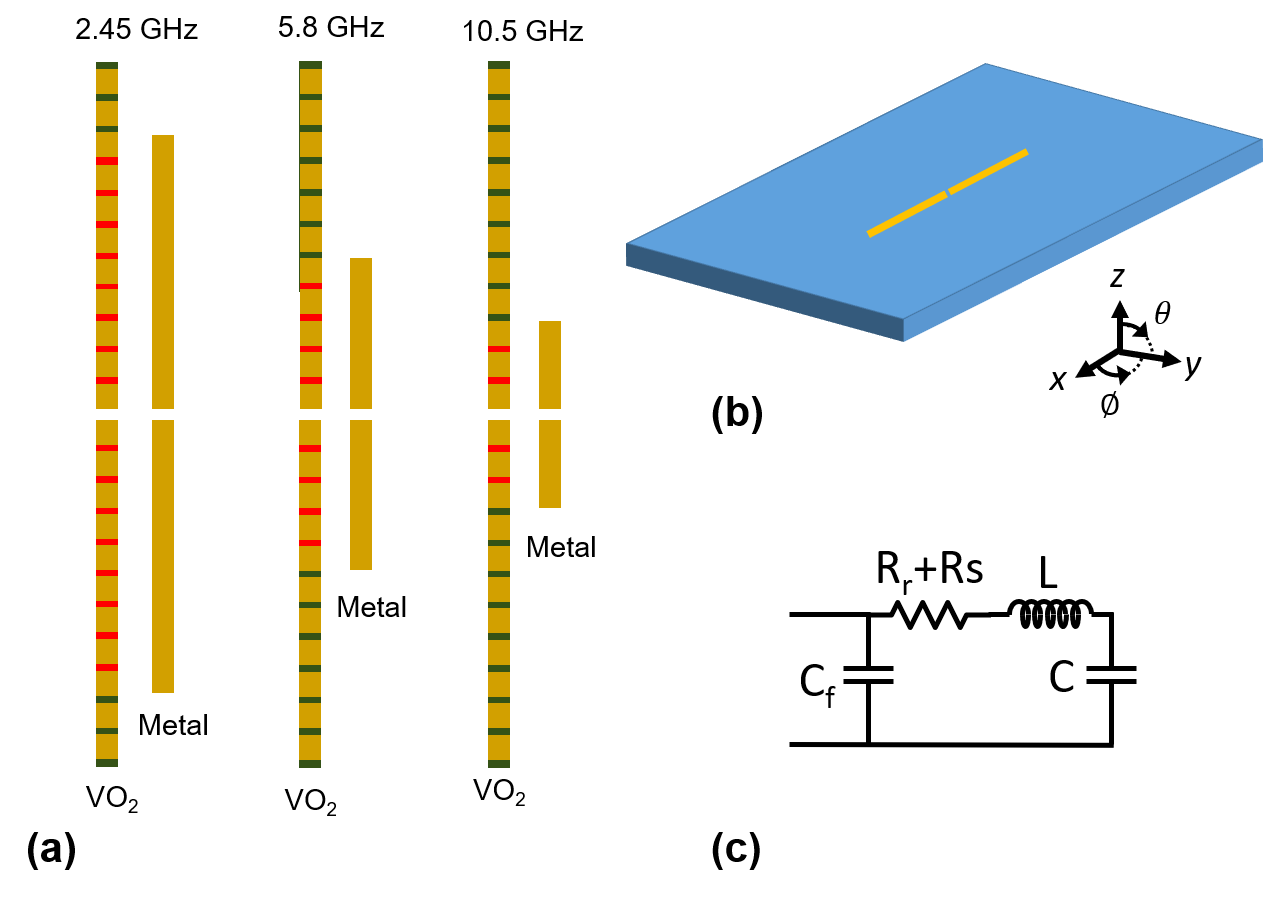}
    \caption{\rev{(a) Simulation setup of three all-metal metal dipoles and three dipoles with {\VO}-metal surface-inclusions at 2.45, 5.8, and 10.5 GHz. Unit cell length is 100\,$\mu$m, feed gap is 100\,$\mu$m, and dipole width is 200\,$\mu$m. {\VO} film thickness is 100\,nm and gap is 100\,nm. Neither the dipole dimensions nor the number of unit-cells are to scale. The {\VO} dipoles are loaded with OFF-state unit-cells (dark green). (b) The dipoles were modeled on a sapphire substrate. (c) The circuit model for each dipole consists of a feed capacitance in parallel with a series RLC circuit with values defined in Table \ref{tab:dipoleParams}.}}
    \label{fig:dipoleSetup}
    %\vspace{-0.5cm}
    \end{figure}}
    
\rev{
    \begin{figure*}[t]
    \centering
    \includegraphics[width=\linewidth]{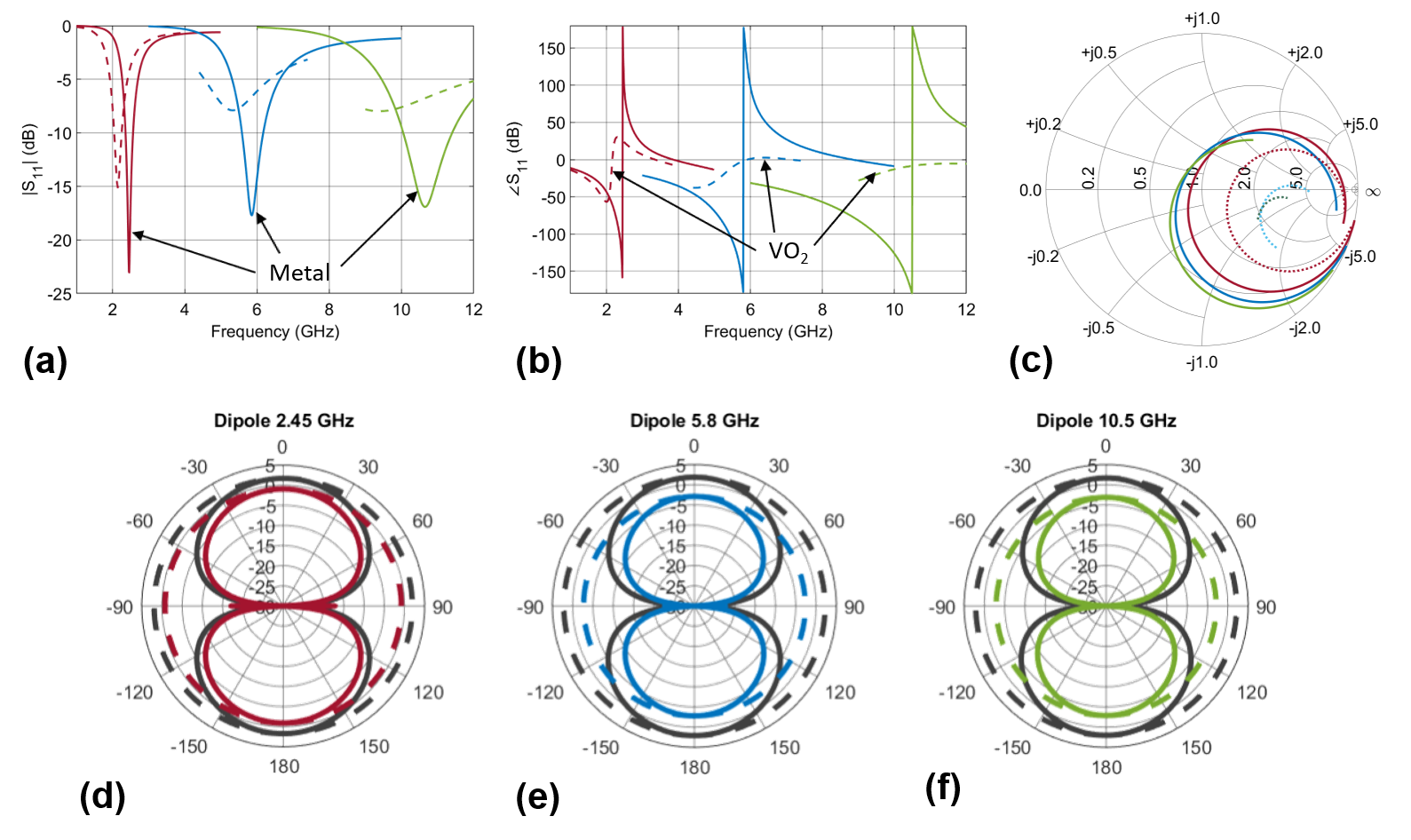}
    \caption{\rev{(a), (b), and (c) plot the magnitude, phase and impedance of each metal (solid) and {\VO} dipole (dashed): red is 2.45 GHz, blue is 5.8\,GHz, green is 10.5\,GHz. d) Radiation patterns are gain patterns for phi cuts 0 (solid) and 90 (dashed). Black is metal and colored is dipole.}}
    \label{fig:dipoleSim}
    %\vspace{-0.5cm}
    \end{figure*}}
    
\rev{As observed in Fig. \ref{fig:fmax} and \ref{fig:ParamLambda}, the suitable frequency range for a 100\,$\mu$m unit-cell is 2 to 14\,GHz (40\,dB isolation), and for a 250\,$\mu$m unit-cell the range is 1 to 12\,GHz (20\,dB isolation). Thus, two practical reconfigurable circuit realizations in these bands are demonstrated in simulation and presented below: a reconfigurable dipole with 100\,$\mu$m unit-cells, and a reconfigurable triple-stub matching network with 250\,$\mu$m unit-cells. These circuits show the substantial benefit of one-dimensional realizations of this concept; a discussion of a two-dimensional implementation circuit, a switch matrix, follows and is accompanied by a comparison with state-of-the-art switches. Demonstration of a two-dimensional reconfigurable circuit is left as future work.}

\rev{\subsection{Reconfigurable Dipole Antenna}}

\rev{Fig.\,\ref{fig:dipoleSetup} shows a simulation model of a reconfigurable {\VO}-metal surface-inclusion dipole from 2.45 to 10.5 GHz. In this section we simulate it and compare its performance with an all-metal versions. The metal dipoles were sized at 38, 14.16, and 7.06\,mm in order to be resonant at 2.45, 5.8, and 10.5\,GHz. The {\VO} dipoles, comprising ON-state unit-cells, are of the same physical length, but also include OFF-state unit-cells at the ends to emulate a realistic reconfigurable dipole. This adds to their physical and electrical length so that they resonate at 2.13, 5.24, and 9.07\,GHz. A {\VO} film thickness of 100\,nm, gap of 100\,nm, and dipole width of 200\,$\mu$m were selected. The dipole was modeled with $\sigma_{\textrm{ON}} = $ 10$^5$\,S/m for ON-state unit-cells and $\sigma_{\textrm{OFF}} = $ 80\,S/m for OFF-state unit-cells. The simulation results are shown in Fig. \ref{fig:dipoleSim}. In order to focus on the performance of the {\VO} dipole, it was excited directly at the feed and not with a transmission line; the feeding element is left as an independent task.}

\rev{
    \begin{figure*}[b]
    \centering
    \includegraphics[width=0.95\textwidth]{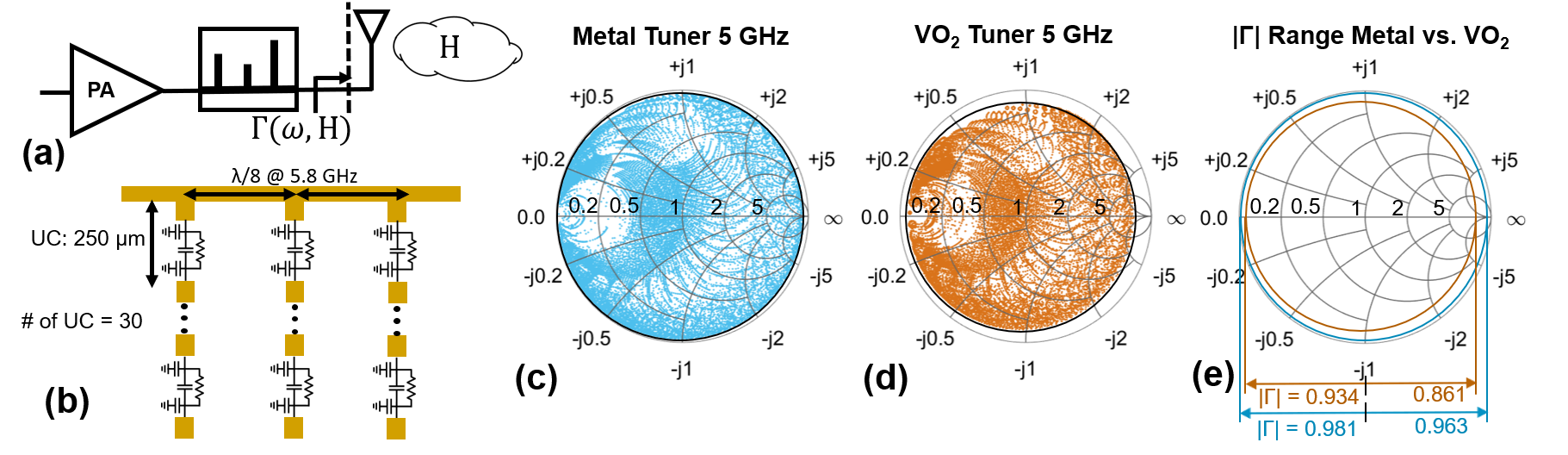}
    \caption{\rev{(a) A {\VO} reconfigurable triple-stub matching network has the potential respond real-time to an antenna's input impedance, which changes not only with frequency but also with the near-field environment, H. (b) The tunable triple stub matching network was simulated with the circuit models extracted from EM simulations of thick {\VO} and microstrip transmission lines. All possible combinations of thirty unit-cells per stub, with a unit-cell length of 250\,$\mu$m were simulated and presented in (d). (c) shows the equivalent for a tunable metallic-only triple-stub matching network. (e) compares the maximum $|\Gamma|$ for a purely metallic stub.}}
    \label{fig:tunerSetup}
    %\vspace{-0.5cm}
    \end{figure*}}
    
\rev{In general the {\VO} dipoles exhibit reasonable match and a dipole radiation pattern at each frequency but they resonate with a lower Q, at a slightly lower frequency, with less gain, and higher input impedance than their metal counterparts. Figure\,\ref{fig:dipoleSim}(a) and (b) show the shift in frequency and damping of the resonance. The shift is attributed to the increased capacitance at the ends of the dipole due to the OFF-state unit-cells, giving the {\VO} dipole a longer electrical length. The damped resonance is attributed to the increase in conductor loss from the presence of {\VO}. In addition to being damped, the resonance of the 5.8 and 10.5 GHz {\VO} dipoles are also shifted capacitively, as seen by the shift towards a negative angle in the resonance's center point, and the rotation on the Smith chart (Fig. \ref{fig:dipoleSim}(c)) about a more capacitive impedance point. This higher and more capacitive input impedance is indicative of a change in the current at the input terminals, which is influenced by both the OFF-state unit-cells terminating the dipole as well as the gap capacitance and {\VO} resistance of each ON-state unit-cell.}

\rev{
    \begin{table}[t]
\renewcommand{\arraystretch}{0.8}
\setlength{\tabcolsep}{5pt}
\caption{Simulated metal and {\VO} dipole antenna and circuit parameters}
\label{tab:dipoleParams}
\centering
\begin{tabular}{rcccccccc}
    \toprule
    Dipole & D\textsubscript{o} & G\textsubscript{o} & $\eta_{cd}$ & R\textsubscript{r}& R\textsubscript{s} & L & C & C\textsubscript{f}\\
    \vhtable (GHz) & (dBi) & (dBi) & & ($\Omega$) & ($\Omega$) & (nH) & (fF) & (fF)\\
    \midrule
    \vhtable    Metal 2.45 & 2.09 & 1.71 & 0.917 & 39.6 & 3.6 & 23.3 & 183 & 137 \\
    \vhtable    Metal 5.8 & 2.19 & 2.07 & 0.973 & 34.9 & 0.96 & 6.78 & 114 & 137 \\
    \vhtable    Metal 10.5 & 2.09 & 2.00 & 0.979 & 28.8 & 0.62 & 2.94 & 81.7 & 137 \\
    \vhtable    {\VO} 2.13 & -0.25 & -0.93 & 0.855 & 60.5 & 10.3 & 26.5 & 210 & 151 \\
    \vhtable    {\VO} 5.24 & 2.25 & -2.81 & 0.312 & 36.7 & 80.7 & 6.69 & 138 & 89.7 \\
    \vhtable    {\VO} 9.07 & 2.01 & -2.90 & 0.322 & 35.7 & 75 & 2.73 & 113 & 68.3 \\
    \bottomrule
%    \vspace*{+1pt}
\end{tabular}
    \vspace*{-1pt}
\end{table}}
    
\rev{The circuit model fit (Fig. \ref{fig:dipoleSetup}(c)) bears the above trends out as well and is documented in Table \ref{tab:dipoleParams}. Specifically, both the conductor resistance R\textsubscript{s} and the series C significantly increase as compared to the metal dipoles. The $1/\sqrt{LC}$ frequency where the {\VO} dipole resonances actually occur are 2.13, 5.24, and 9.07 GHz. The smaller the {\VO} dipole, the greater the shift for two reasons: the OFF-state unit-cells have less isolation with higher frequency, and their physical extent comprises a larger fraction of the total length for shorter dipoles.}
    
\rev{The {\VO} dipole gain patterns, efficiencies, radiation resistance R\textsubscript{r} and the conductor resistance R\textsubscript{s} are calculated at 2.13, 5.24, and 9.07 GHz. The gain patterns are identical to the metal patterns in shape, as seen in Fig. \ref{fig:dipoleSim}(d), (e), and (f).  The maximum gain is -0.93, -2.81, and -2.9 dBi for each {\VO} dipole, respectively. Since the substrate is sapphire, all losses are attributed to conductor losses. While the gain of these dipoles is lower than an isotropic radiator, this reconfigurable dipole has a 4.25 tuning ratio or 124$\%$ bandwidth, rendering it incredibly useful for a wide range of frequencies. Table \ref{tab:antennas_art} compares this approach with several other reconfigurable antenna technologies.}

\rev{\subsection{Reconfigurable Triple-Stub Matching Network}}

\rev{In addition to the reconfigurable dipole, we have simulated the performance of a triple stub matching network utilizing the circuit models developed from EM simulations of T\textsubscript{VO2} $= 100$\,nm and gap size of G $= 100$\,nm from 10\,MHz to 50\,GHz. Because a triple stub tuner is capable of transforming a source impedance to any region of the Smith chart, this is an excellent one-dimensional application of our PCM-metal surface-inclusion concept. We chose to use the circuit model because (i) this allows us to simulate all possible stub length combinations over a wide band incredibly more efficiently than with EM simulations, (ii) the unit-cell length can easily be varied, and (iii) this demonstrates the usefulness of the circuit models for rapidly prototyping new distributed circuits.}

\begin{table}[t]
\renewcommand{\arraystretch}{0.7}
\setlength{\tabcolsep}{3pt}
\caption{Comparison of Reconfigurable Antennas}
\label{tab:antennas_art}
\centering
\begin{tabular}{rccccc}
    \toprule
    Tech & Freq. & Tune & Eff.\,/\,Gain & Speed & DC Bias\\    
    \vhtable  & (GHz) & (\%) & (\%)\,/\,(dBi) & ($\mu$sec) & \\
    \midrule
    \vhtable    Liquid \cite{shiroma_liquidmetal_monopole} & 4.48-5.52 & 20.8 & 90.4\,/\,0.3 & Slow & N/A\\
    \vhtable    Origami \cite{origami} & 1.78-3.64 & 68.6 & 12.7\,dBi & Deployable & N/A\\
    \vhtable    Ferrite \cite{ferrite} & 9.75-10.7 & 9.29 & 86\,/\,6.5 & -- & 200\,kA/m\\
    \vhtable    Piezoelec.  \cite{piezoelectric} & 2.78-2.98 & 9 & 8.3\,dBi & -- & 200\,V\\
    \vhtable    Optoelec. \cite{opto_monopole} & 2.65-10.3 & 118 & -- & -- & 590\,mW\\
    \vhtable    Sim \VO \cite{Gerislioglu_VO2AntennaWithMicroheaterArray_2017} & 2-10 & 133 & -27.5 to -12\,dBi & 80\,msec & 121.5\,$\mu$W\\
    \vhtable    GeTe \cite{GeTe_antenna} & 24.4-29 & 17.2 & $>$\,73\,/\,$>$\,3.02 & ~0.03 & --\\
    \vhtable    \rev{This Work & 2.13-9.07 & 124 & 0.312\,/\,-2.81 & -- & --}\\
    \bottomrule
%    \vspace*{+1pt}
\end{tabular}
    \vspace*{-1pt}
\end{table}
%\end{minipage}

\rev{
    \begin{figure}[ht]
    \centering
    \includegraphics[width=\columnwidth]{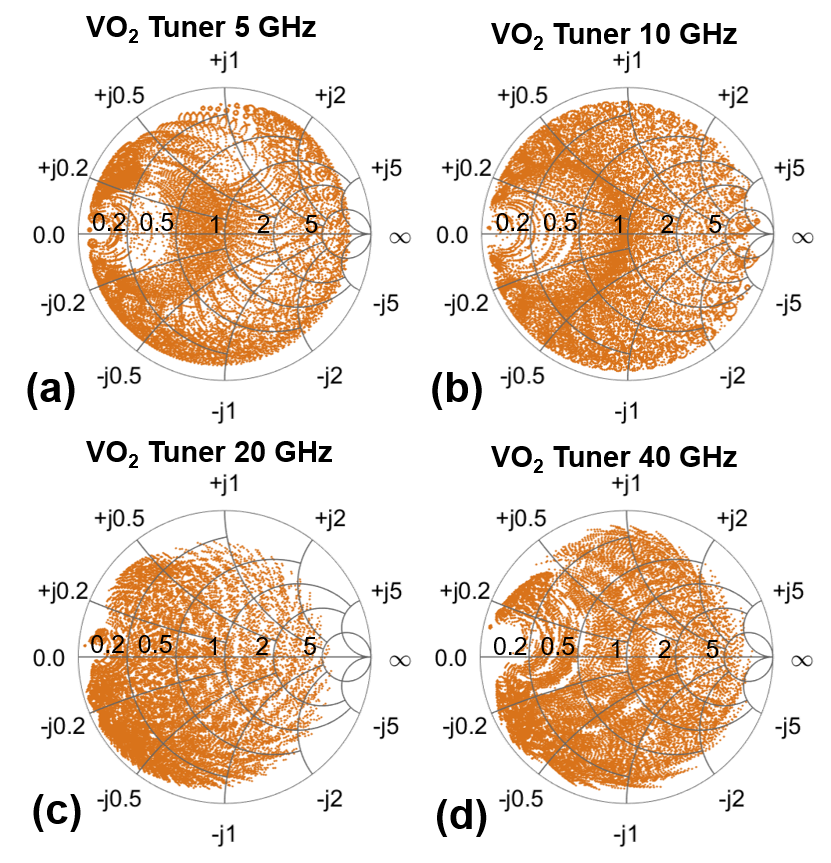}
    \caption{\rev{The {\VO} tunable triple-stub matching network simulated at 5, 10, 20 and 40 GHz.}}
    \label{fig:tunerFreq}
    %\vspace{-0.5cm}
    \end{figure}}
    
\rev{A possible scenario for a triple-stub tuner can be seen in Fig. \ref{fig:tunerSetup}(a), in which a tuner is used to match a power amplifier to an antenna in a dynamic near-field environment. The antenna environment causes significant and possibly rapidly varying changes in the antenna input impedance, which necessitates real-time adjustment to the antenna matching network. This is useful in e.g., mobile phones where antennas must operate in isolation as well as when brought near a human body or placed on various surfaces. Figure\,\ref{fig:tunerSetup}(b) shows the simulation setup, where a microstrip triple-stub tuner on a 100\,$\mu$m sapphire substrate has stubs that are capable of being reconfigured by selecting the ON or OFF state circuit model. A metal thickness of 2\,$\mu$m was used. The tuner was modeled with $\sigma_{\textrm{ON}} = $ 10$^5$\,S/m for ON-state unit-cells and $\sigma_{\textrm{OFF}} = $ 5\,S/m for OFF-state unit-cells. The unit-cell size selected was 250\,$\mu$m, and the total number of unit-cells available for tuning on each stub is 30, for a possible 30$^3$ stub length combinations with only 30$\times3=90$ control heaters (consuming an estimated 36\,$\mu$W of power in the heater array). The spacing between each stub is $1/8$ at 5.8 GHz.}

\rev{Figure \ref{fig:tunerSetup}(c) and (d) show the region covered on the Smith chart for a purely metallic tuner and a reconfigurable {\VO}-metal surface-inclusion stub at 5 GHz. While the {\VO} tuning network comes close to the performance of the metallic counterpart, the ON-state loss and OFF-state isolation in the {\VO} reduces slightly the $|\Gamma|$ of the network. Figure \ref{fig:tunerSetup}(e) compares the maximum $|\Gamma|$ of each tuning network: $|\Gamma|$\textsubscript{VO2} = 0.934, $|\Gamma|$\textsubscript{M} = 0.981 at low impedances and $|\Gamma|$\textsubscript{VO2} = 0.861, $|\Gamma|$\textsubscript{M} = 0.963 at high impedances.}

\rev{Figure \ref{fig:tunerFreq} shows the performance of the {\VO} matching network at 5, 10, 20, and 40\,GHz. The network is still able to capture the majority of the Smith chart with very good granularity considering a large unit-cell size of 250\,$\mu$m. At 40 GHz, where 250\,$\mu$m is 31$^\circ$ of a guided wavelength, the stubs become periodic in $\lambda/2$ after 11 unit-cells. The matching network, however, can still present a different impedance each time the stub's physical length becomes periodic in $\lambda/2$ because of the increase in total loss of the stub. Therefore, this triple-stub matching network has an 8:1 tunable bandwidth.}

\rev{\subsection{Reconfigurable Switch Matrix}}

\rev{The above one-dimensional demonstrations of reconfigurable circuits testify to the efficacy of this approach by achieving significant reconfigurability within a single application. To obtain full programmability of distributed circuits, a two-dimensional reconfigurable surface is needed. The performance of this reconfigurable surface is fundamentally linked to the primitive's (unit-cell) Figure of Merit (FOM), which means that, even for the best of switches, a reconfigurable surface will never achieve the same performance as a dedicated metallic circuit. Rather, the benefit of a reconfigurable surface with high FOM primitives is that a circuit can be tuned or reconfigured in near-real-time to meet changes in requirements or changes in a radio's environment.}

\rev{While {\VO} has a high FOM comparable to state-of-the-art switches (see Table \ref{tab:state-of-art}) and can be controlled by thermal heaters that are electrically isolated from the EM circuit, this concept is not tied to a single switch technology. MEMS switches have the highest FOM but suffer from long-term reliability and high activation voltages. Additionally, designing bias lines for large switch matrices that maintain low EM coupling becomes an extremely challenging task. Germanium telluride (GeTe) has a comparable FOM to {\VO}, but has the advantage of zero static power dissipation with electrical triggering \cite{GeTe_SPST}. Furthermore, a PCM film-based approach has the advantage of preserving the mode continuity between unit-cells much more so than a FET, MEMS, or diode would preserve it.}

\begin{table}[t]
\renewcommand{\arraystretch}{0.7}
\setlength{\tabcolsep}{3pt}
\caption{Comparison with Switch Technologies}
\label{tab:state-of-art}
\centering
\begin{tabular}{rccccc}
    \toprule
    Technology & FOM & Loss/Isol. & Freq. & Power & Speed \\% & Area \\
    \vhtable & (THz) & (dB) & (GHz) & & ($\mu$sec)\\
    \midrule
    \vhtable    {\VO} \cite{VO2_SPST} & 26.5 &  $<$0.5/$>$30 & DC-50 & 37.5 mW & 0.025 \\
    \vhtable    GeTe \cite{GeTe_SPST} & 7.8 & $<$ 0.3/$>$20 & 0-26 & No static & 0.15-2 \\
    \vhtable    FET \cite{FET_CMOS} & 0.263 & -- & 0.5-10 & 60\,uW & 34.2 \\
    \vhtable    MEMS \cite{MEMS_SPNT} & 16.5 & $<$\,0.26\,/\,$>$\,21 & DC-12 & 63 VDC & 7-14 \\
    \vhtable    MEMS \cite{MEMS_Comparison} & 63.7 & R\textsubscript{ON}/C\textsubscript{OFF}=$0.22\Omega,11$fF & 0.2-4 & --- & --- \\
    \vhtable    This work & 1.5 & $<$\,0.867\,/\,$>$\,10.5 & DC-10 & -- & -- \\
    \vhtable    This work & 36.79 & $<$\,0.015\,/\,$>$\,6.7 & DC-10 & $<36\,\mu$W & -- \\
    \vhtable    {\VO} 35\,nm & 10.4 & R\textsubscript{ON},C\textsubscript{OFF}=$3\Omega,5$fF & 0.01-50 & -- & -- \\
    \bottomrule
\end{tabular}
\end{table}

\section{Conclusion} \label{sec:conclusion}

We have characterized, for the first time, transmission lines made of thin {\VO} films with metallic inclusions and have found that thicker films (100--200\,nm) and lower percentage of {\VO} ($\ll$\,5\,\%) are required to realize low-loss reconfigurable distributed circuits. While this blend of PCM material and metal is the first step in enabling low-loss, fully programmable distributed circuits, the trade-off is a reduction in current confinement that degrades with frequency. We have also successfully modeled these structures with EM simulations, and have extracted both a transmission line model at 50\,GHz and a lumped-element circuit model of the {\VO} gaps that is valid from 0.01--50\,GHz. \rev{We have concluded with a demonstration of a reconfigurable dipole at 2.14, 5.24, and 9.07\,GHz, and a tunable matching network with high-$|\Gamma|$ at 5, 10, 20, and 40\,GHz.} Arbitrary 2D designs will benefit from our EM model parameters for full-wave simulations, whereas 1D structures can be rapidly prototyped using the transmission line and circuit model parameters. \srev{Reconfigurable antennas, matching networks, and filters are a straightforward extension of this work when realized along a single dimension (e.g., wire antennas and stub matching networks).}

Future work to expand this concept to 2D will require studying the characteristic impedance of poorly defined conductors as well as investigating and comparing the performance of different waveguides. Microstrip, for example, is likely to perform better than planar waveguides, such as slotlines and CPW, given that the entire ground plane can remain metal and that the OFF-state {\VO} will not increase shunt conductance between signal and ground.

\section{Acknowledgements}
The authors gratefully acknowledge help from Professor R. Engel-Herbert at Pennsylvania State University for supplying the {\VO} samples. The authors would also like to thank Edit Varga for assistance with AFM measurements, and Nicholas Estes for helpful conversations about {\VO}.

\ifCLASSOPTIONcaptionsoff
  \newpage
\fi

\bibliographystyle{IEEEtran}
\bibliography{MTT_ReconfigurableRF}

%\newpage
\begin{IEEEbiography}[{\includegraphics[width=1in,height=1.25in,clip,keepaspectratio]{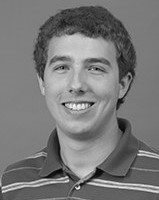}}]{David A. Connelly} (Student Member, IEEE) is currently a Ph.D student in Electrical Engineering at the University of Notre Dame in Indiana, U.S.A. In 2016, he graduated with a B.S., Electrical Concentration from LeTourneau University in Longview, Texas, U.S.A. His research interests are in exploiting novel materials, including phase change materials, and magnetic spin-waves for RF signal processing and control.
\end{IEEEbiography}

\begin{IEEEbiography}[{\includegraphics[width=1in,height=1.25in,clip,keepaspectratio]{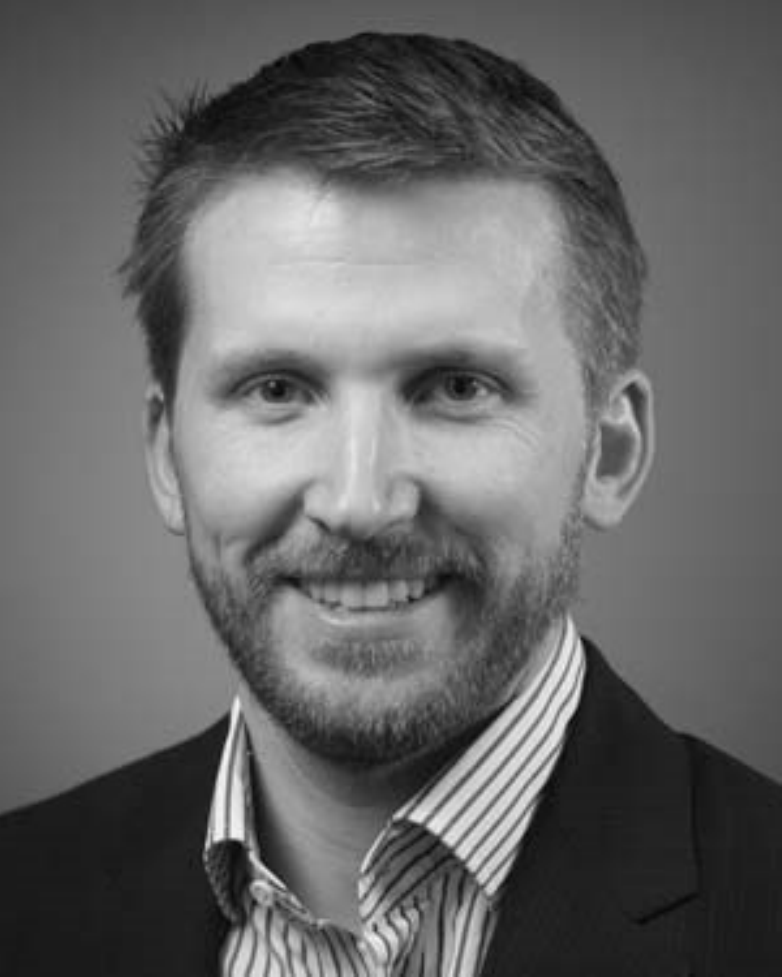}}]{Jonathan~D.~Chisum} (S'02--M'06--SM'17) received the Ph.D. in Electrical Engineering from the University of Colorado at Boulder in Boulder, Colorado USA, in 2011. 

From 2012 to 2015 he was a Member of Technical Staff at the Massachusetts Institute of Technology Lincoln Laboratory in the Wideband Communications and Spectrum Operations groups. His work at Lincoln Laboratory focused on millimeter-wave phased arrays, antennas, and transceiver design for electronic warfare applications. In 2015 he joined the faculty of the University of Notre Dame where he is currently an Assistant Professor of Electrical Engineering. His research interests include millimeter-wave communications and spectrum sensing with an emphasis on low-power and low-cost technologies. His group focuses on gradient index (GRIN) lenses for low-power millimeter-wave beam-steering antennas, nonlinear (1-bit) radio architectures for highly efficient communications and sensing up through millimeter-waves, as well as reconfigurable RF circuits for wideband distributed circuits and antennas. 

Dr. Chisum is a senior member of the IEEE, a member of the American Physical Society, and an elected Member of the U.S. National Committee (USNC) of the International Union or Radio Science's (URSI) Commission D (electronics and photonics). He is the past Secretary and current Vice-chair for USNC URSI Commission D: Electronics and Photonics.

\end{IEEEbiography}

% that's all folks
\end{document}